\newcommand{\teff}{T$_{\mathrm{eff}}$}
\newcommand{\Teff}{T$_{\mathrm{eff}}$}
\newcommand{\logg}{log $g$}

\newcommand{\kms}{km~s$^{-1}$}

\newcommand{\vmicro}{$\rm{v_{micro}}$}

\newcommand{\project}[1]{\textsl{#1}}
\newcommand{\gaia}{\project{Gaia}}

\documentclass[a4paper,fuseAMS,leqn,usenatbib]{mnras}

\usepackage{graphicx}
\usepackage{amssymb}
\usepackage{amsmath}
\usepackage{times}
\usepackage{subfigure}
\usepackage[T1]{fontenc}
\usepackage{ae,aecompl}
\usepackage{textcomp}

\usepackage{hyperref}

\title[Chemistry of Metal-Poor Bulge Stars]{The COMBS survey I: Chemical Origins of Metal-Poor Stars in the Galactic Bulge\thanks{Based on observations collected at the European Southern Observatory under ESO programme: 089.B-069 }}

 \author[Lucey, et al. 2019]{Madeline~Lucey$^{1}$\thanks{E-mail:m\_lucey@utexas.edu}, Keith~Hawkins$^{1}$, Melissa~Ness$^{2,3}$, Martin~Asplund$^{5,6}$, \newauthor Thomas~Bensby$^{4}$,  Luca~Casagrande$^{5}$, Sofia~Feltzing$^{4}$, Kenneth~C.~Freeman$^{5}$, \newauthor  Chiaki~Kobayashi$^{7}$ and Anna~F.~Marino$^{8,9}$
\\
$^{1}$Department of Astronomy, The University of Texas at Austin, 2515 Speedway Boulevard, Austin, TX 78712, USA \\
$^{2}$Center for Computational Astrophysics, Flatiron Institute,
162 5th Ave., New York, NY 10010, USA\\
$^{3}$Department of Astronomy, Columbia University, 550 W 120th St., New York, NY, 10027, USA \\
$^{4}$Lund Observatory, Department of Astronomy and Theoretical Physics, Box 43, SE-221\,00 Lund, Sweden \\
$^{5}$Research School of Astronomy and Astrophysics, The Australian National University, Canberra, ACT 2611, Australia \\
$^{6}$ARC Centre of Excellence for All Sky Astrophysics in 3 Dimensions (ASTRO 3D) \\
$^{7}$ Centre for Astrophysics Research, School of Physics, Astronomy and Mathematics, University of Hertfordshire, Hatfield AL10 9AB, UK \\
$^{8}$Dipartimento di Fisica e Astronomia 'Galileo Galilei' --- Univ. di
Padova, Vicolo dell’Osservatorio 3, Padova, IT-35122, Italy \\
$^{9}$Centro di Ateneo di Studi e Attivita' Spaziali 'Giuseppe Colombo' ---
CISAS, Via Venezia 15, Padova, IT-35131, Italy \\}

\date{Accepted 2019 July 1. Received 2019 May 31; in original form 2019 March 26}

\pubyear{2019}

\begin{document}
\label{firstpage}
\pagerange{\pageref{firstpage}--\pageref{lastpage}}
\maketitle

\begin{abstract}
  Chemistry and kinematic studies can determine the origins of stellar population across the Milky Way. The metallicity distribution function of the bulge indicates that it comprises multiple populations, the more metal-poor end of which is particularly poorly understood. It is currently unknown if metal-poor bulge stars ([Fe/H] $<$ -1 dex) are part of the stellar halo in the inner most region, or a distinct bulge population or a combination of these. Cosmological simulations also indicate that the metal-poor bulge stars may be the oldest stars in the Galaxy. In this study, we successfully target metal-poor bulge stars selected using SkyMapper photometry. We determine the stellar parameters of 26 stars and their elemental abundances for 22 elements using R$\sim$ 47,000 VLT/UVES spectra and contrast their elemental properties with that of other Galactic stellar populations. We find that the elemental abundances we derive for our metal-poor bulge stars have lower overall scatter than typically found in the halo. This indicates that these stars may be a distinct population confined to the bulge. If these stars are, alternatively, part of the inner-most distribution of the halo, this indicates that the halo is more chemically homogeneous at small Galactic radii than at large radii. We also find two stars whose chemistry is consistent with second-generation globular cluster stars. This paper is the first part of the Chemical Origins of Metal-poor Bulge Stars (COMBS) survey that will chemo-dynamically characterize the metal-poor bulge population.
\end{abstract}


\section{Introduction}

\label{sec:Introduction}
Understanding galaxy formation and evolution is now a realizable objective of astrophysics given the ensemble of data and tools in hand. Bulges are major components of most spiral galaxies \citep[e.g.,][]{Gadotti2009}. However, it is not well understood how they form and evolve. By studying the large number of resolved stars in our own Galactic bulge, we can gain new insight into the formation and evolution of bulges. However, historically, this has been difficult. The level of crowding in the bulge makes it difficult to resolve individual stars without very large telescopes. In addition, high levels of dust extinction towards the Galactic center cause dimming, making it hard to achieve high signal-to-noise ratios for observations of resolved bulge stars. 

There is now many observations of the Galactic bulge from a multitude of surveys. Imaging surveys such as the Optical Gravitational Lensing Experiment \citep[OGLE,][]{Udalski2002}, the Two Micron All-Sky Survey \citep[2MASS,][]{Skrutskie2006}, and Vista Variables in the Via Lactea survey \citep[VVV,][]{Minniti2010,Saito2012a} have used red clump giant stars (RCGs) to reveal an X-shaped structure \citep{Nataf2010,McWilliam2010,Saito2012b}. This was possible because RCGs can be used as standard candles \citep{Stanek1998,Hawkins2017}. Spectroscopic surveys such as the Bulge Radial Velocity Assay \citep[BRAVA,][]{Rich2007}, the Abundances and Radial velocity Galactic Origins \citep[ARGOS,][]{Freeman2013}, the GIRAFFE Inner Bulge Survey \citep[GIBS,][]{Zoccali2014}, and the HERMES Bulge Survey \citep[HERBS,][]{Duong2019} have measured the radial velocities and chemical abundances of bulge stars.  \citet{Ness2012} found that only stars with [Fe/H]\footnote{Chemical abundances are reported in the standard way, as a logarithmic ratio with respect to solar values. Mathematically, [X/Y] = $\rm{log}\left(\frac{N_X}{N_Y}\right)_{star}-\rm{log}\left(\frac{N_X}{N_Y}\right)_\odot$ where $N_X$ and $N_Y$ are the number of each element X and Y per unit volume, respectively.} $>$ -0.5 dex participate in the B/P structure. On the other hand, stars with lower [Fe/H] have been shown to have distinct kinematics and morphological structure  \citep{Sharples1990,Rich1990,Zhao1994,Soto2007,Hill2011,Ness2016,Zoccali2017}.

The metallicity distribution function (MDF) of the bulge provides further evidence for multiple populations. Using 14,150 stars in the bulge, \citet{Ness2013} found the MDF to have five distinct components with peaks at metallicities of about  +0.15, -0.25, -0.7, -1.18, -1.7 dex. They associate these peaks with the B/P bulge (+0.15 and -0.25 dex peaks), the thick disk (-0.7 dex peak), the metal-weak thick disk (-1.18 dex peak) and the stellar halo (-1.7 dex peak). The three higher metallicity peaks dominant with only about 5\% of stars with metallicities < -1.0 dex \citep{Ness2016}. Other studies have found similar results, demonstrating the MDF of the bulge has multiple components \citep[e.g.,][]{Zoccali2008,Johnson2013a,Zoccali2017,Bensby2013,Rojas-Arriagada2014,Bensby2017,Rojas-Arriagada2017,Duong2019}.

The low metallicity end of the MDF has recently become of interest. State of the art simulations have shown low-mass Population III stars could still exist today \citep[e.g.,][]{Clark2011,Greif2012,Bromm2013}. It has become increasingly clear that if these Population III stars exist in our galaxy, they will be found in the central regions \citep{White2000,Brook2007,Diemand2008}. Further work has shown the metal-poor stars in the bulge are more likely to be older than equally metal-poor stars located elsewhere in the Galaxy \citep{Salvadori2010, Tumlinson2010}.

Given the predicted initial mass function (IMF) from simulations of metal-free star formation, it is thought that a significant fraction of the first stars would explode as pair-instability supernovae (PISNe) \citep{Heger2010}. Simulated yields from PISNe show over 90\% stars primarily enriched from PISNe and formed in atomic cooling halos have metallicities around $\sim$ -2.5 dex \citep{Karlsson2008}. Given that most of the oldest stars are thought to have formed ex-situ (e.g., in atomic cooling halos) and end up in the center of the Galaxy \citep{Nakasato2003,El-Badry2018}, it is possible that the oldest stars in the Galaxy are in the bulge with metallicities $\leq$ -2 dex  \citep{Chiappini2011,Wise2012,Cescutti2018}. The progenitors in which the oldest stars formed are too faint to be detected, even with the James Webb Space Telescope \citep{Gardner2006}. So these Galactic stars, concentrated to the inner regions, provide the only window into the formation and evolution of these small galaxies at high redshift. We note that prior
efforts to study the most metal-poor and first stars have largely focused on the Galactic halo, and dwarf satellites, far from the centre of the Galaxy \citep[e.g.,][]{Frebel2006,Norris2007,Christlieb2008,Keller2014,Starkenburg2017}.

The discovery of low mass ($\sim$ 0.7 $M_{\sun}$) stars from a first-star population could provide a vital constraint on the initial mass function (IMF) in the first galaxies. Although this would not give insight into the IMF for metal-free stars it would be very relevant to the evolution of the earliest stars to form in the universe. Additionally, this potentially oldest stars population can test different models of early enrichment. 

There has been a recent effort to search for the most metal-poor and Population III stars in the bulge. These searches have made significant progress despite the large distance, crowding and high dust extinction in the bulge. The bulge is also the most metal-rich component of the Galaxy, leaving only 1 in 20 stars to have [Fe/H] <-1 dex  \citep{Fulbright2006,Ness2016}. Although they cannot definitively determine if their target stars are located in the bulge,  \citet{GarciaPerez2013} used infrared spectroscopy of $\sim$ 2,400 stars toward the bulge and found five stars with -2.1 dex $\leq$ [Fe/H] $\leq$ -1.6 dex. \citet{Schlaufman2014} found three stars with -3.0 dex < [Fe/H] < -2.7 dex. The Extremely Metal-poor BuLge stars with AAOMega spectroscopic survey \citep[EMBLA,][]{Howes2014} was the first survey to successfully target metal-poor bulge stars. \citet{Howes2014} found four bulge stars with -2.72 dex $\leq$ [Fe/H] $\leq$ -2.48 dex and \citet{Howes2015} found 23 bulge stars with [Fe/H] <-2.3 dex with the most metal-poor star at [Fe/H] = -3.94 dex. Finally, \citet{Howes2016} added 10 more stars with -3.0 dex $\leq$ [Fe/H] $\leq$ -1.6 dex. \citet{Koch2016} analyzed 3 Bulge stars within 4 kpc of the Galactic center with -2.56 dex $\leq$ [Fe/H] $\leq$ -2.31 dex. In total, there are on the order of 50 studied metal-poor stars in the bulge.

It is important to note that these studies of metal-poor bulge stars could be contaminated. In other words, it is yet to be determined if the detected metal-poor stars in the bulge are truly the oldest stars or if they have other origins. For example, it is possible that these stars are simply halo stars with eccentric orbits that pass through the bulge. \citet{Howes2015} measured the orbits of 10 metal-poor bulge stars and found only seven of the stars to have tightly-bound bulge-like orbits. Another possible origin scenario is accreted material from a dwarf galaxy such as Gaia-Enceladus \citep{Belokurov2018,Helmi2008} or massive disrupted globular clusters \citep{Kruijssen2015,Shapiro2010,Bournaud2016}. \citet{Siqueira-Mello2016} found 3 of the 5 metal-poor bulge stars they studied had chemical abundances similar to the metal-poor bulge globular clusters, NGC 6522 and M62.  These stars could be from protogalactic clusters \citep[e.g.,][]{Diemand2005,Moore2006} and therefore still some of the oldest stars in the Galaxy. 

The goal of this paper is to explore the chemistry of the most metal-poor stars in the Galactic bulge in order to determine their origin. In particular, we want to search for clues as to if these stars are distinct from the Milky Way populations of the thick disk and stellar halo, which have well described chemical properties \citep[e.g.,][]{Reddy2003,Adibekyan2012,Yong2013,Roederer2014,Bensby2014,Battistini2015,Battistini2016}.  Chemical markers that would differentiate the oldest stars from stars of the Galactic halo include sodium, aluminum, copper and manganese, which are expected to be much lower in the oldest stars given the metallicity dependence of the yields \citep{Kobayashi2011c}. If a star is predominantly enriched from PISNe, as some of the oldest stars are thought to be, it would have almost no elements heavier than Fe \citep[e.g.,][]{Karlsson2008,Kobayashi2011b,Takahashi2018}. Given the first stars have a top-heavy IMF \citep{Tumlinson2006,Bromm2013}, the stars enriched from the first stars would have higher levels of $\alpha$-enhancement than the thick disk or halo. The theoretical yields of $\alpha$ elements from a non-rotating PISNe are on the order of [$\alpha$/Fe] $\sim$ 2 dex  \citep{Takahashi2018} while the lowest metallicity stars in the local disk have [$\alpha$/Fe] $\sim$ 0.4 dex.

We present the discovery of 22 metal-poor bulge stars and additional analysis of 4 ARGOS stars. In total, we perform abundance analysis of 26 stars for 22 elements. In section \ref{sec:selection} we describe the selection of metal-poor bulge stars and in section \ref{sec:data} we describe the observations. As described in section \ref{sec:analysis}, the data is reduced using standard techniques and the FLAMES/UVES reduction pipelines. For the stellar parameter and abundance analysis we use the Brussels Automatic Code for Characterizing High accUracy Spectra \citep[BACCHUS,][]{Masseron2016} which is further described in section \ref{sec:analysis}. Finally, in section \ref{sec:results} we present and discuss the results.

\section{Selection of Metal-Poor Bulge Stars} \label{sec:selection}

\begin{table*}

\caption{ARGOS targets with the ARGOS stellar parameters}
\label{tab:argos}
\begin{tabular}{ccccccc}
\hline\hline
2MASS ID & ID & RA & DEC & $\rm{T_{eff}}$ & log(g) & [Fe/H] \\
& & (deg) & (deg) & (K) & &  \\
\hline
J18240990-3341561 & 7383.0 & 276.04140 & -33.69890 & 5179 & 2.50 & -2.22 \\
J18182580-3739409 & 25782.0 & 274.60750 & -37.66138 & 5038& 2.05 & -2.56 \\
J18550481-1949206 & 12931.0 & 283.77004 & -19.82239 & 5296 & 2.64 & -2.36 \\
J18531035-2050078 & 5262.0 & 283.29310 & -20.83540 & 5193 & 2.86 & -2.46 \\
J18153438-2727353 & 42011.0 & 273.89320 & -27.45980 & 5393 & 3.15 & -2.46 \\
\hline
\end{tabular}
\flushleft{Column 1 and 2 gives the 2MASS ID and the ID from this study, respectively. Column 3 and 4 gives the coordinates of these targets. The stellar parameters (\teff, \logg, and [Fe/H], respectively) determined for these stars in the ARGOS survey are given columns 5, 6, and 7.}
\end{table*}

\begin{figure}
    \centering
    \includegraphics[width=\columnwidth]{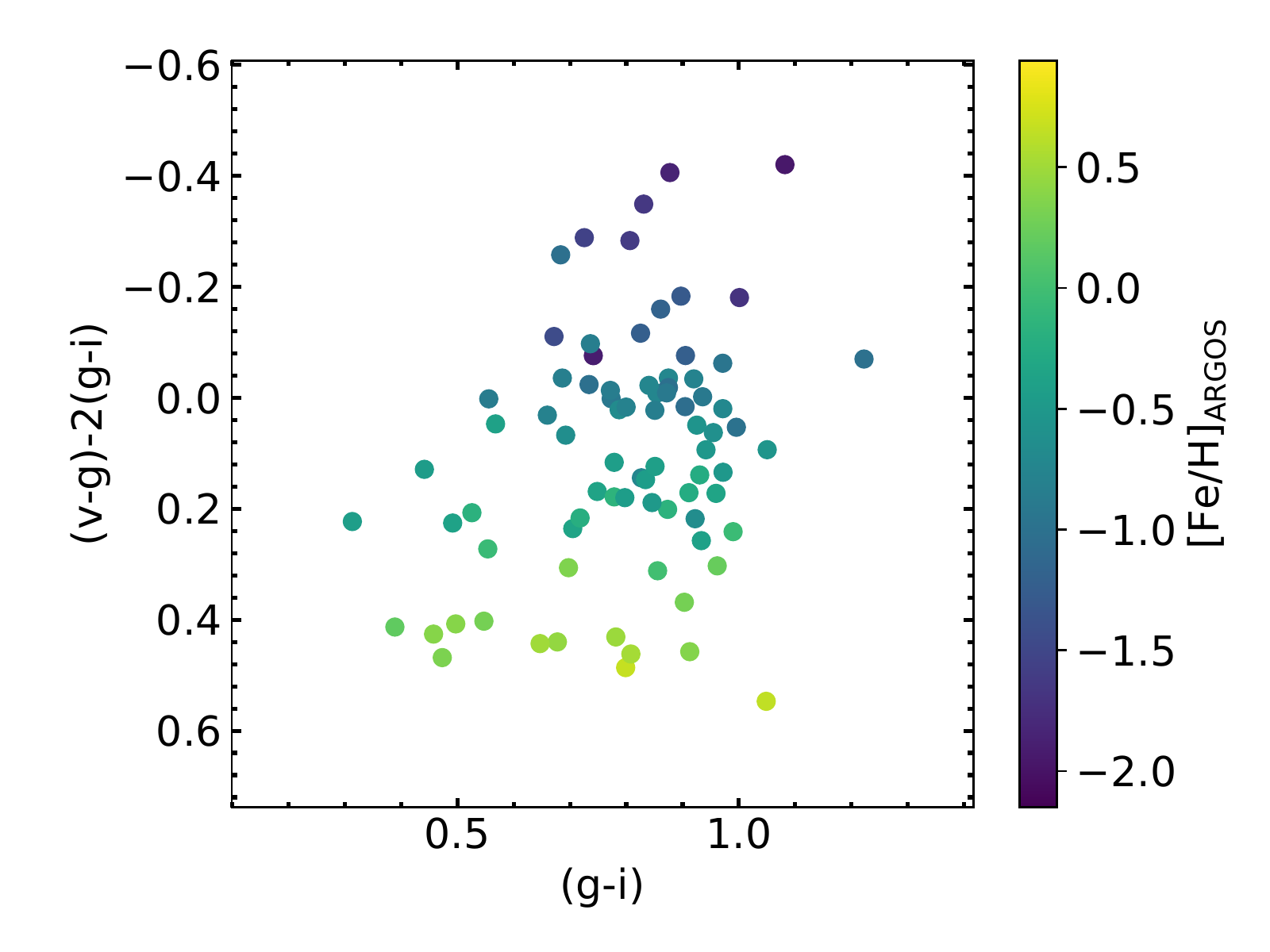}
    \caption{Shows the relationship between SkyMapper colors and metallicity. The x-axis is (g-i) photometry and the y-axis (v-g)-2(g-i). Each point is colored by its spectroscopically derived metallicity from the ARGOS survey. The most metal-poor stars have the most negative (v-g)-2(g-i) colour index and largest (g-i) values.} 
    \label{fig:phot_met}
\end{figure}

To date, it has been extremely difficult to select the most metal-poor stars in the Galactic bulge region, due to crowding and high extinction in the centre of the Galaxy and the large fraction of relatively metal-rich stars in the bulge.  The metal-poor population represents only a tiny fraction of the overall stellar population in the inner Galaxy. The combination of large spectroscopic surveys such as ARGOS, GIBS and APOGEE \citep{Freeman2013,Zoccali2014,Majewski2017}, which each observe up to tens of thousands of bulge stars, and which determine an [Fe/H] measurement for each star, and the photometric SkyMapper survey, which provides metallicity sensitive colours for orders of magnitude more stars across the Galaxy and into the bulge, are absolutely essential to pre-select metal-poor candidates in order to make progress.

Our program represents a specific targeted set of observations of metal-poor stars with [Fe/H] < -2.0 dex and lying within 3.5 kpc of the Galactic centre. We used a combination of ARGOS spectroscopic and SkyMapper photometry to make our target selection. ARGOS measured stellar parameters for about 28,000 stars in the inner regions of the Galaxy. The ARGOS fields span latitudes of $b  = -5^\circ, -7.5^\circ$ and $-10^\circ$ and longitudes extending out into the thin and thick disks of $l = +26^\circ$ to $-31^\circ$. From the ARGOS medium resolution (R=$\lambda/\Delta \lambda \sim$  11,000) spectra across the Ca-triplet region  stellar parameters, \teff, \logg, [Fe/H], $\rm{[\alpha/Fe]}$ were determined, and distances were calculated for all stars. In total, the metal-poor ARGOS sample includes 17 stars with -2.8 dex $\leq$ [Fe/H] $\leq$ -2.2 dex with Galactocentric radii between 1 and 3 kpc. From these 17 primary ARGOS targets (14.5 < V < 17.5), we selected for our high resolution UVES observations, 5 ARGOS targets with magnitudes 14.5 < V < 15.5, which ARGOS measured metallicities between -2.5 to -2.25 ($\pm$ 0.15) dex and $\alpha$ enhancement all at $\sim$ 0.7 ($\pm 0.15$) dex and within 3.5 kpc from the Galactic center.

These primary 5 targets are listed in Table \ref{tab:argos}. These targets are supplemented with a larger sample of metal-poor stars selected from our SkyMapper\footnote{Located at Siding Spring Observatory in Australia, SkyMapper is a 1.35m automated wide-field survey telescope with the goal of mapping the entire southern sky down to $\sim$ 20-21 mag with photometry in six filters, \textit{uv} (unique to SkyMapper) and \textit{griz}  \citep[Sloan Digital Sky Survey like,][]{Fukugita1996}. For further description of the SkyMapper photometric system we refer the reader to \citet{Bessell2011}.} photometry within each 25 arcminute UVES/FLAMES field \citep[similarly to][]{Howes2014}, with SkyMapper photometry calibrated using the ARGOS [Fe/H] determinations. This approach represents a highly efficient survey for metal-poor stars within the inner Milky Way.  As outlined in \citet{Howes2014},  the SkyMapper survey features a filter set optimized for stellar astrophysics. In particular it provides excellent resolution of stellar metallicity \citep{Keller2007}: the rare metal-poor stars can be isolated from the bulk of bulge stars by their UV excess with a low level of contamination. The photometric selection was made, however, using the preliminary (and uncalibrated) commissioning SkyMapper photometry. Individual cuts were made on each field using the colour sensitivity as shown in Figure \ref{fig:phot_met}. The use of commissioning data suffices for our purpose of identifying metal-poor stars. Current SkyMapper data has shown to perform well at mapping stellar metallicities \citep{Casagrande2019} although this does not include regions close to the Galactic plane where the current pipeline is not optimized to deal with high stellar crowding \citep{Wolf2018}.


Our UVES/FLAMES fields are at longitudes of $0^\circ, +5^\circ, +15^\circ$ and $-10^\circ$;  all within the region predicted by simulations to be populated with the highest density of the oldest stars in the Galaxy. We used the photometric sensitivity to make a selection of targets to fill all FLAMES and remaining UVES fibers (beyond our 5 primary targets), and have a sample of a total of 40 UVES (R$\sim$ 47,000) stars and 640 FLAMES stars (R $\sim$ 20,000) and in this work, we examine our high resolution UVES targets.



We required between 3 to 9.5 hours on each primary ARGOS target to obtain a signal-to-noise, S/N $\sim$ 50. This SNR requirement is linked to the requirement to reach a precision $< $0.2 dex in our elemental abundance measurements, which is sufficient to distinguish between different stellar populations. 


\section{Data} \label{sec:data}

\subsection{UVES Spectroscopic Data}
\label{sec:uves}
\begin{table*}

\caption{Observational Properties of our Target Stars}
\label{tab:obs}
\begin{tabular}{cccccccccccc}
\hline\hline
Star & 2MASS ID & RA & DEC & l & b & $G$ & $\rm{N_{spec}}$ & RV & $\rm{RV_{scatter}}$& $\rm{SNR_{L}}$ \\

& & (deg) & (deg) & (deg) & (deg) & (mag) & & (km $\rm{s^{-1}}$) & (km $\rm{s^{-1}}$) & ($\rm{pixel^{-1}}$) \\
\hline
644.0 & 18241657-3332426 & 276.06904 & -33.54519 & -0.03 & -9.45 & 13.85 & 20 & 19.4$\pm$0.2 & 0.53 & 129 \\
697.0 & 18242041-3327187 & 276.08508 & -33.45521 & 0.05 & -9.43 & 15.29 & 20 & -274.7$\pm$0.3 & 0.36 & 54 \\
1067.0 & 18245245-3343393 & 276.21858 & -33.72760 & -0.14 & -9.65 & 15.51 & 19 & -48.8$\pm$0.1 & 0.42 & 34 \\
1490.0 & 18251658-3339277 & 276.31912 & -33.65771 & -0.04 & -9.69 & 15.20 & 20 & 20.7$\pm$0.2 & 0.49 & 56 \\
1670.0 & 18252843-3336055 & 276.36850 & -33.60153 & 0.03 & -9.70 & 15.35 & 20 & -39.6$\pm$0.2 & 0.49 & 43 \\
1697.2 & 18521929-2047049 & 283.08042 & -20.78472 & 14.36 & -9.50 & 15.49 & 7 & 65.2$\pm$0.2 & 0.59 & 23 \\
2700.0 & 18183598-3735190 & 274.64996 & -37.58862 & -4.23 & -10.22 & 15.59 & 8 & -23.8$\pm$0.4 & 0.36 & 25 \\
2860.0 & 18524481-2045130 & 283.18675 & -20.75362 & 14.43 & -9.57 & 13.71 & 7 & -123.0$\pm$0.4 & 0.59 & 92 \\
3083.0 & 18545773-2009145 & 283.74054 & -20.15403 & 15.20 & -9.79 & 13.94 & 8 & 199.1$\pm$0.2 & 0.08 & 55 \\
3230.3 & 18532702-2044083 & 283.36262 & -20.73566 & 14.52 & -9.72 & 15.62 & 7 & -54.2$\pm$0.1 & 0.63 & 25 \\
3655.0 & 18154293-2742578 & 273.92888 & -27.71607 & 4.34 & -5.15 & 12.77 & 15 & 56.2$\pm$0.3 & 0.28 & 12 \\
4239.1 & 18521121-2041371 & 283.04675 & -20.69365 & 14.43 & -9.43 & 15.97 & 5 & 196.5$\pm$0.2 & 0.47 & 15 \\
4475.0 & 18543151-2005517 & 283.63129 & -20.09771 & 15.21 & -9.67 & 14.71 & 12 & 21.6$\pm$0.2 & 0.78 & 49 \\
4648.0 & 18175515-3732343 & 274.47983 & -37.54287 & -4.25 & -10.08 & 15.36 & 8 & -100.6$\pm$0.3 & 0.11 & 21 \\
4953.1 & 18524113-2040183 & 283.17142 & -20.67176 & 14.50 & -9.53 & 15.08 & 7 & -108.4$\pm$0.4 & 0.53 & 35 \\
5126.3 & 18190264-3730527 & 274.76100 & -37.51466 & -4.12 & -10.27 & 14.49 & 8 & 50.5$\pm$0.1 & 0.23 & 44 \\
5199.0 & 18143907-2734132 & 273.66283 & -27.57035 & 4.36 & -4.87 & 15.18 & 1 & 230.1$\pm$0.1 & 0.00 & 5 \\
5529.0 & 18154601-2735429 & 273.94171 & -27.59527 & 4.46 & -5.10 & 15.07 & 19 & 143.6$\pm$0.1 & 0.22 & 34 \\
5780.0 & 18150819-2736489 & 273.78417 & -27.61361 & 4.37 & -4.99 & 13.66 & 15 & -98.2$\pm$0.1 & 0.19 & 6 \\
5953.0 & 18144441-2737321 & 273.68504 & -27.62559 & 4.32 & -4.92 & 14.65 & 18 & -39.5$\pm$0.1 & 0.16 & 16 \\
6373.1 & 18523066-2037237 & 283.12775 & -20.62327 & 14.52 & -9.47 & 15.91 & 12 & -103.9$\pm$0.4 & 0.95 & 24 \\
6382.0 & 18143785-2726551 & 273.65775 & -27.44866 & 4.47 & -4.81 & 15.11 & 18 & 106.2$\pm$0.1 & 0.20 & 46 \\
6531.3 & 18531383-2037240 & 283.30767 & -20.62335 & 14.60 & -9.62 & 14.73 & 6 & 66.3$\pm$0.2 & 0.12 & 31 \\
6577.0 & 18550952-2001104 & 283.78967 & -20.01956 & 15.35 & -9.77 & 14.96 & 8 & -28.4$\pm$0.1 & 0.05 & 41 \\
6805.0 & 18150652-2728214 & 273.77717 & -27.47264 & 4.50 & -4.92 & 14.25 & 19 & -126.4$\pm$0.1 & 0.17 & 24 \\
7064.3 & 18181536-3729226 & 274.56404 & -37.48963 & -4.17 & -10.12 & 14.38 & 8 & -77.3$\pm$0.2 & 0.33 & 42 \\
7362.0 & 18541908-1959101 & 283.57954 & -19.98616 & 15.29 & -9.58 & 16.20 & 11 & -7.6$\pm$0.2 & 0.54 & 24 \\
7604.0 & 18184012-3728409 & 274.66717 & -37.47805 & -4.12 & -10.18 & 14.63 & 8 & -79.1$\pm$0.2 & 0.44 & 30 \\
9071.0 & 18191186-3724179 & 274.79942 & -37.40498 & -4.01 & -10.25 & 15.59 & 1 & 41.1$\pm$0.3 & 0.00 & 4 \\
9094.0 & 18551367-1955180 & 283.80700 & -19.92169 & 15.44 & -9.74 & 14.84 & 8 & -54.8$\pm$0.3 & 0.05 & 49 \\
9761.0 & 18543714-1953441 & 283.65479 & -19.89560 & 15.41 & -9.60 & 15.13 & 12 & 86.9$\pm$0.2 & 0.56 & 41 \\
11609.0 & 18250368-3333070 & 276.26533 & -33.55195 & 0.03 & -9.60 & 15.02 & 20 & 114.1$\pm$0.1 & 0.51 & 62 \\
12909.0 & 18251962-3328046 & 276.33179 & -33.46797 & 0.14 & -9.62 & 15.60 & 20 & 100.0$\pm$0.4 & 0.48 & 35 \\
12931.0 & 18550481-1949206 & 283.77004 & -19.82239 & 15.52 & -9.67 & 14.72 & 12 & 88.3$\pm$0.9 & 0.41 & 50 \\
25782.0 & 18182580-3739409 & 274.60750 & -37.66138 & -4.31 & -10.22 & 14.85 & 8 & -58.7$\pm$0.6 & 0.26 & 35 \\
42011.0 & 18153438-2727353 & 273.89329 & -27.45983 & 4.56 & -5.00 & 15.12 & 19 & 70.8$\pm$0.6 & 0.56 & 41 \\

\hline
\end{tabular}
\flushleft{
Column 1 gives the identifier in this survey, while column 2 gives the 2MASS identifier. Columns 3, 4, 5 and 6 give the coordinates in right ascension, declination, Galactic longitude and Galactic latitude, respectively. The \gaia\ $G$-band magnitude is given in column 7. The number of co-added spectra is tabulated in column 8. The mean measured RV of those spectra is given in column 8 and the scatter between spectra is given in column 10. Finally, the SNR for the lower/blue chip is given in column 11.  }

\end{table*}

Spectroscopic data for the 40 bulge targets, selected as described in Section \ref{sec:selection}, were obtained with the UVES instrument on the European Southern Observatory's (ESO) Very Large Telescope (VLT). UVES is a high resolution optical spectrograph with wavelength coverage 3000-11000 \AA. The spectrograph has two arms, the RED arm and the BLUE arm. The BLUE arm is for the ultraviolet wavelengths (3000-5000 \AA) and the RED arm is for the visual wavelengths (4200-11000 \AA). The RED arm has two CCDs, lower/blue and upper/red. Observations for this work were taken in the standard RED580 setup. This setup has a wavelength coverage of 4726-6835 \AA\ with a gap (5804-5817 \AA) between the lower/blue and upper/red chips and R $\sim$ 47,000. For more details about UVES we refer the reader to \citet{Dekker2000}.

The data for this project were taken in the `MOS' mode for the FLAMES/UVES instrument. Raw data can be found within the ESO archive\footnote{\url{http://archive.eso.org/eso/eso_archive_main.html}} (Program ID: 089.B-0694). As noted within the ESO archive, reduced Phase~3 data products are not provided for UVES spectra observed in the `MOS' mode. Therefore, we have reduced the data using version 2.9.1 of the EsoReflex interface\footnote{\url{https://www.eso.org/sci/software/esoreflex/}}. Within the EsoReflex interface we made use of the FLAMES-UVES workflow for the data reduction.   

In short, the EsoReflex package performs a traditional data reduction workflow. Namely, it completes a bias subtraction, fiber order trace, computation  and correction for both the detector pixel-to-pixel gain variations and the blaze function. After which it extracts the spectrum  and performs the wavelength calibration for each fiber. For more details we refer the reader to Section~9 of the UVES-fibre pipeline manual. Descriptions on how to download EsoReflex, its use, and the exact calibration steps we refer the reader to the UVES-FIBRE instrument pipeline package\footnote{\url{ftp://ftp.eso.org/pub/dfs/pipelines/uves/uves-fibre-pipeline-manual-18.11.pdf}}, and the accompanying tutorial\footnote{\url{ftp://ftp.eso.org/pub/dfs/pipelines/uves/uves-fibre-reflex-tutorial-1.11.pdf}}.

\begin{figure}
    \centering
    \includegraphics[width=\columnwidth]{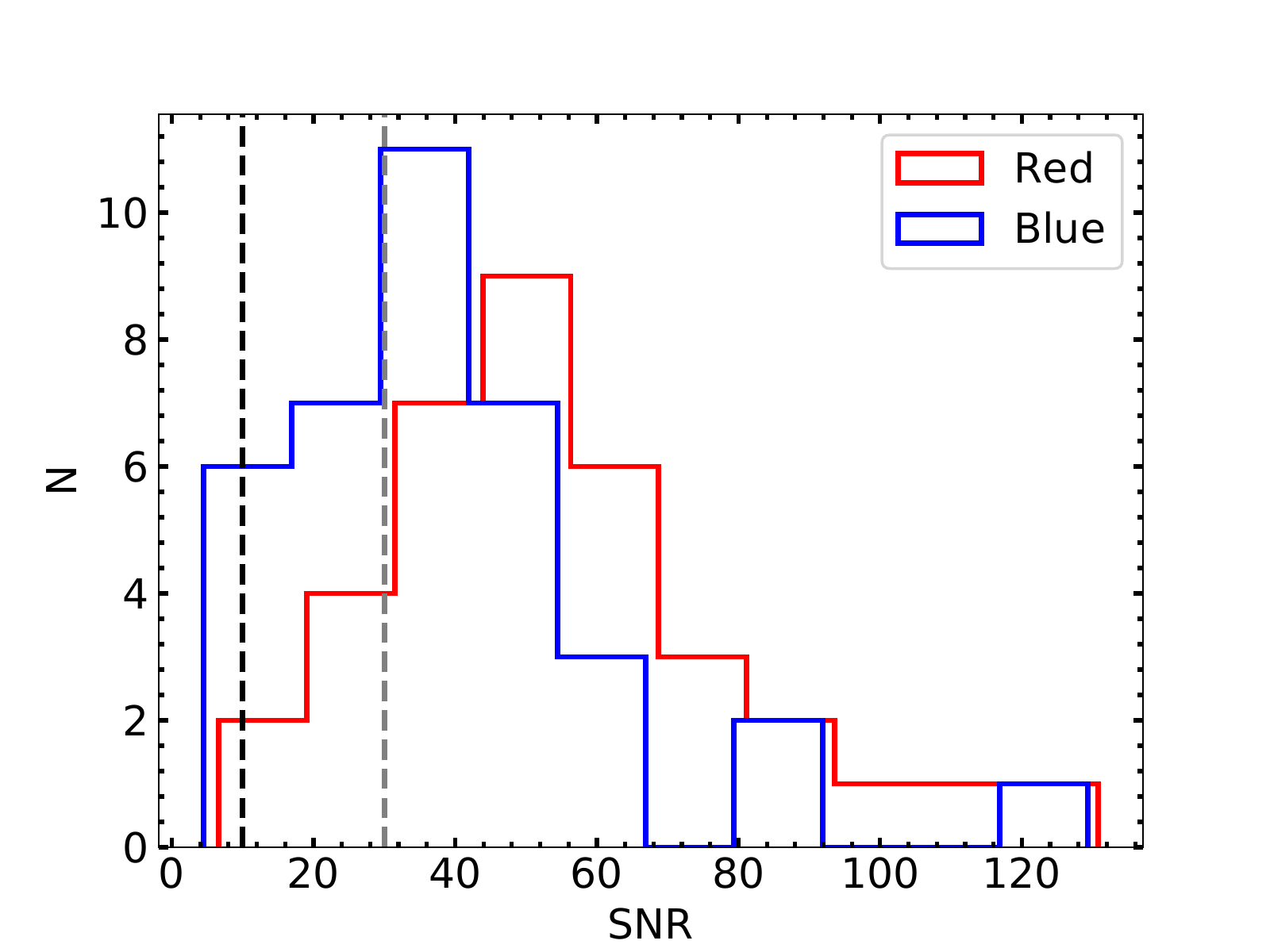}
    \caption{Shows the distribution of the  signal-to-noise ratio (SNR) per pixel in the final bulge sample in both the UVES blue/lower (blue line) and red/upper (red line) chips. The SNR for the blue/lower chip is measured at wavelengths 5353 to 5354.2 \AA, 5449.6 to 5450.49 \AA, and  5464.6 to 5465.4 \AA. The SNR for the red/upper chip is measured at wavelengths 6328.1 to 6329.7 \AA, 6446.7 to 6447.5 \AA, and  6705.5 to 6706.1 \AA.  The black dashed line shows the SNR cut of 10 $\rm{pixel^{-1}}$. The grey dashed line shows the low SNR cut of 30 $\rm{pixel^{-1}}$. Although we report the results, we flag stars with 10 $\rm{pixel^{-1}}$ < SNR < 30 $\rm{pixel^{-1}}$ in our abundance analysis. }
    \label{fig:SNR}
\end{figure}

\begin{figure*}
    \centering
    \includegraphics[width=\linewidth]{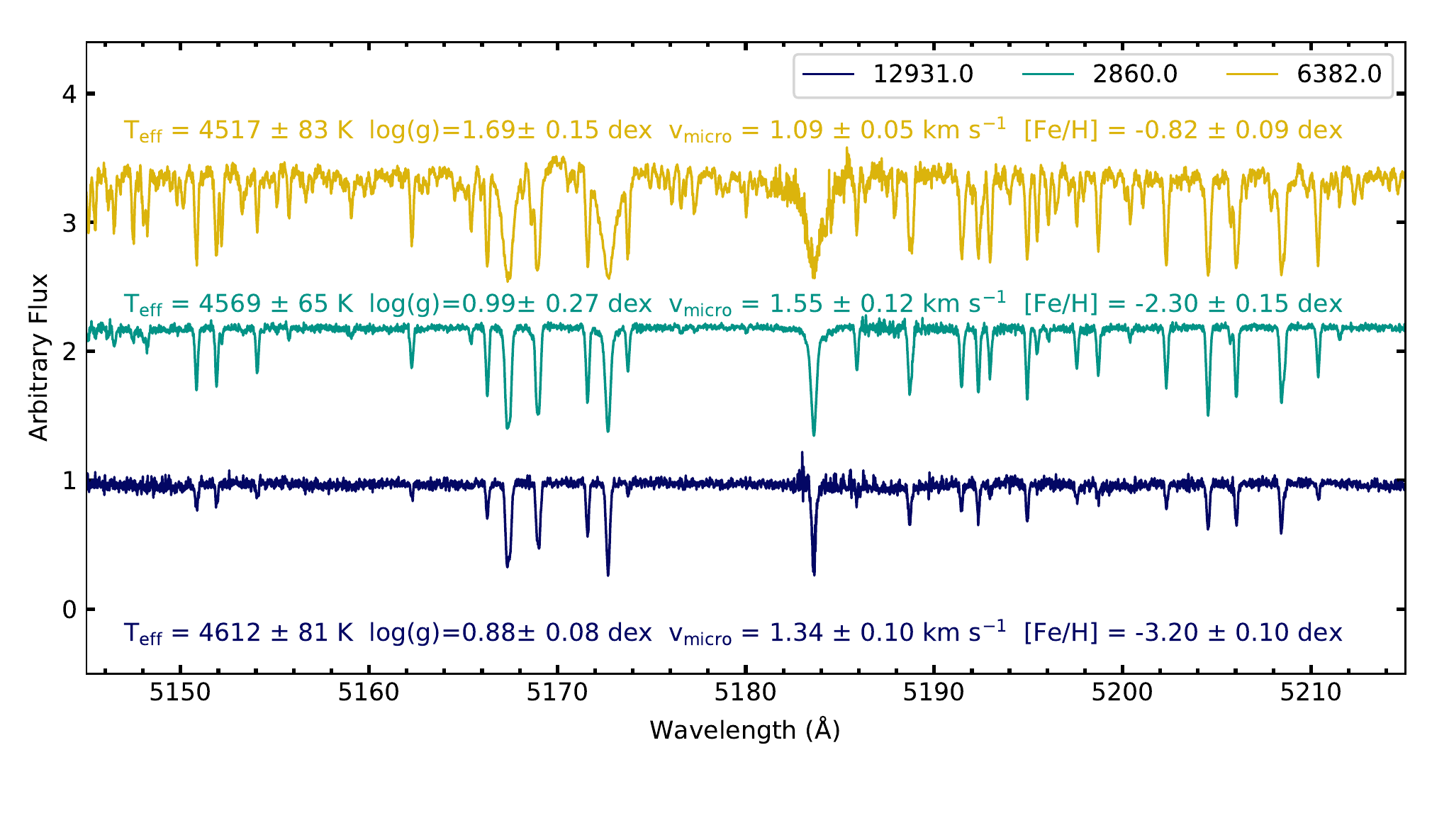}
    \caption{The observed spectra in the Mg triplet region (5145-5215 \AA) of a few target stars, specifically 6577.0 (dark blue solid line), 2860.0 (green solid line), and 6805.0 (yellow solid  line).}
    \label{fig:spec}
\end{figure*}

Each bulge star in this work has been observed multiple times. Therefore, once the data had been reduced, extracted and wavelength calibrated from the EsoReflex interface, we co-added the spectra from unique bulge targets. To do this, we start by using iSpec \citep{Blanco-Cuaresma2014} to fit a  continuum to each spectra using a third order spline. We note that it is significantly easier to find the continuum in metal-poor stars. Once the continuum is fit, the spectra are radial velocity (RV) corrected. RVs are determined using a cross-correlation with respect to Arcturus. Finally, spectra for the same object which had signal-to-noise ratio (SNR) larger than $>$10 $\rm{pixel^{-1}}$  were co-added. One star is removed from the analysis because none of its spectra had SNR > 10 $\rm{pixel^{-1}}$. Stars were rejected if the scatter in radial velocity between the individual visits was larger than a few \kms. This is done because it is not clear if the wrong star made its way into the fiber or if the star has radial velocity variability. This criteria removed three stars. We also removed another 3 stars that have a final SNR in the red/lower chip < 10 $\rm{pixel^{-1}}$. The SNR for the blue/lower chip is measured at wavelengths 5353 to 5354.2 \AA, 5449.6 to 5450.49 \AA, and  5464.6 to 5465.4 \AA\ while the SNR for the red/upper chip is measured at wavelengths 6328.1 to 6329.7 \AA, 6446.7 to 6447.5 \AA, and  6705.5 to 6706.1 \AA. We have a final sample of 33 stars for spectral analysis. The distribution of the final SNR per pixel for the final sample in both the blue/lower and red/upper chip can be found in Fig.~\ref{fig:SNR}. We note that the typical SNR for spectra in this study is SNR $\sim$50 $\rm{pixel^{-1}}$ in the red chip. For reference, we also show in Fig.~\ref{fig:spec} the final reduced, extracted 1-D, continuum and RV corrected spectra for 3 bulge targets. These spectra are what will be used to derive stellar parameters and elemental abundances. 

\subsection{Gaia}

In this study, we use \gaia\ DR2 data to confirm that our target stars are located in the bulge. Combining the Galactic coordinates of our stars with distances derived from \gaia\ DR2 parallaxes, gives us the location of our stars with respect to the Galactic center. However, most of our stars have  fractional parallax errors $>$ 20\% in \gaia\ DR2. Therefore, estimating the distance by inverting the parallax will give unreliable results. The distance catalog from \citet{Bailer-Jones2018} uses more sophisticated methods, namely Bayesian inference with a weak distance prior, to accurately take the large parallax errors into account when determining the distances. Therefore, these distances are more reliable than a simple inverted parallax method and we use them when determining the Galactic positions of the stars in our, and other, samples. 

\begin{figure}
    \centering
    \includegraphics[width=\columnwidth]{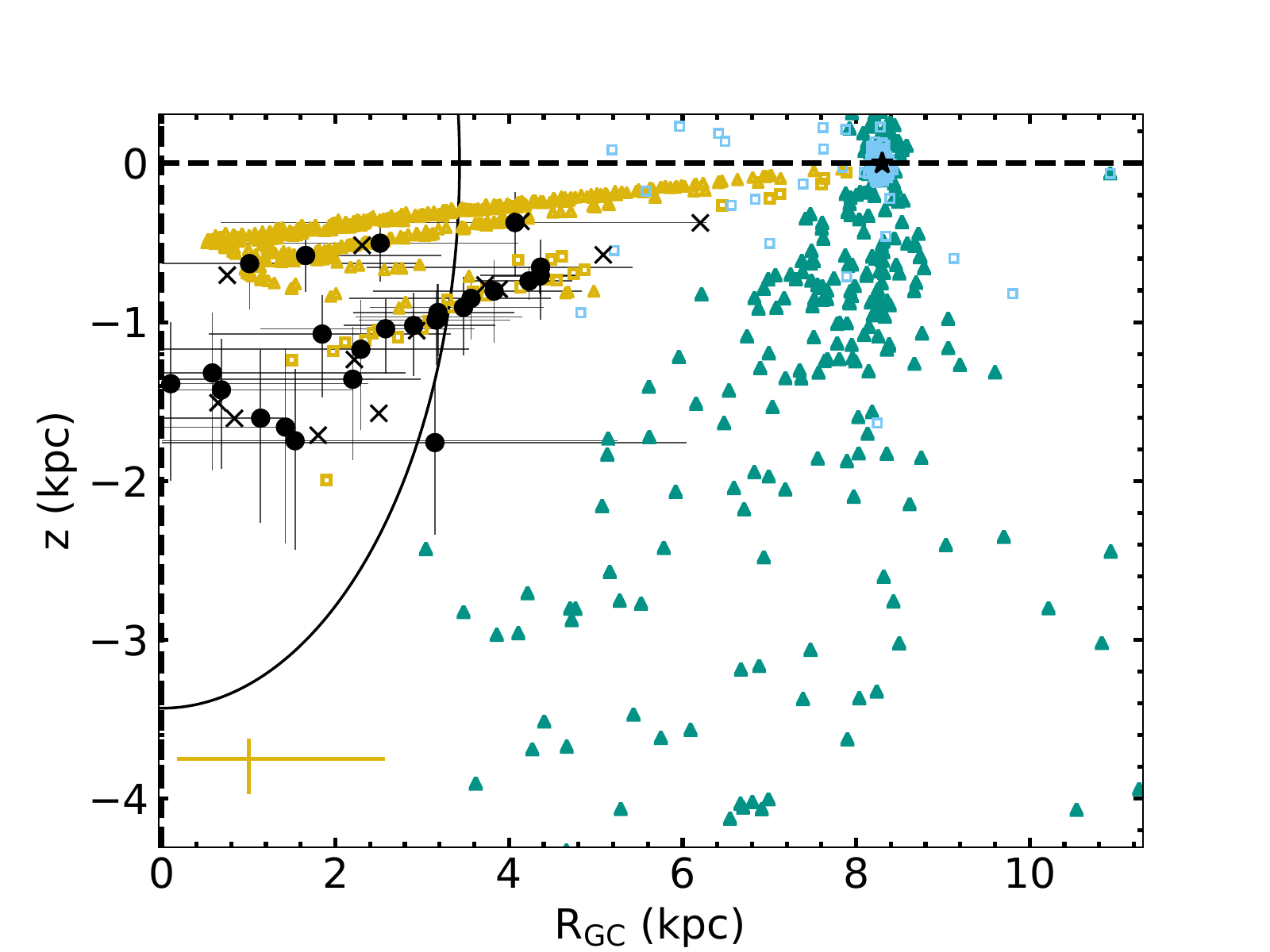}
    \caption{Demonstrating the Galactic position of our target bulge stars compared to a few of the literature samples used in this study. Stars for which we successfully derived abundances are shown as black filled circles, while stars we observed but for which we do not report abundances are shown as black xs. Previous bulge studies are shown in yellow, halo studies are shown in green and disk studies are shown in light blue. The bulge studies included in this figure are \citet[][yellow open squares]{Howes2016}  and \citet[][yellow triangles]{Gonzalez2015}. For comparison, a median error bar for these studies is shown in the yellow in the bottom right corner. The disk study shown is \citet[][light blue open squares]{Bensby2014}  and the halo study is \citet[][green triangles]{Roederer2014}. Distances for each sample are taken from the \citet{Bailer-Jones2018} distance catalog. The Sun is shown as a black star. The solid  line denotes a spherical radius of 3.43 kpc. The typical error in $\rm{R_{GC}}$ for our sample is 2.57 kpc while the typical error in z is 0.40 kpc.}
    \label{fig:rz}
\end{figure}
The distances from \citet{Bailer-Jones2018} combined with the Galactic coordinates show that our sample is comprised of bulge stars. Figure \ref{fig:rz} shows the Galactocentric radius of our target stars. The error bars are calculated by determining the Galactocentric radius for lower and upper bound of the 68 \% confidence interval given in \citet{Bailer-Jones2018}. In addition, we show a local disk study  \\ \citep[][light blue open squares]{Bensby2014}, a halo study \citep[][green triangles]{Roederer2014} and two bulge studies \citep[][yellow open squares and yellow triangles, respectively]{Howes2016,Gonzalez2015}. The distances for each study are taken from the distance catalog in \citet{Bailer-Jones2018}.  Compared to the local disk study and the halo study our stars are much closer to the Galactic center. As shown in Figure \ref{fig:rz}, the samples from both \citet{Howes2016} and \citet{Gonzalez2015} are contaminated with stars with a Galactocentric radius > 5 kpc. The majority of our stars are within 3.43 kpc which \citet{Robin2012} defines as the simplest criteria for a bulge star. Despite the large parallax errors in \gaia\ DR2 for our sample, we find that the Galactocentric radius of our sample is similar to \citet{Howes2016}, which is expected given the similar selection method. Therefore, we conclude our sample is spatially consistent with bulge stars and may be more representative of the bulge than the previous studies. We note that the Galactocentric velocity distribution is consistent with previous bulge studies \citet[e.g.,][]{Howard2008}. In the next part of this survey, we will perform a detailed study of the kinematics of these stars to determine if they are confined to the bulge.

\section{Stellar Parameter and Elemental Abundance Analysis} \label{sec:analysis}

\par Stellar parameters and elemental abundance analysis was done using Brussels Automatic Code for Characterizing High accUracy Spectra \citep[BACCHUS,][]{Masseron2016}. The current released version generates synthetic spectra using the MARCS model atmosphere grid \citep{Gustafsson2008} and the TURBOSPECTRUM radiative transfer code \citep{Alvarez1998, Plez2012}. Also used are the fifth version of the Gaia-ESO linelist for atomic lines which includes hyperfine structure (Heiter at al., in preparation), and molecular lines for CH \citep{Masseron2014}, CN, NH, OH, MgH, $\rm{C_2}$ (T. Masseron, private communication), SiH (Kurucz linelists), and TiO, ZrO, FeH, CaH (B. Plez, private communication).

\begin{figure}
    \centering
    \includegraphics[width=\columnwidth]{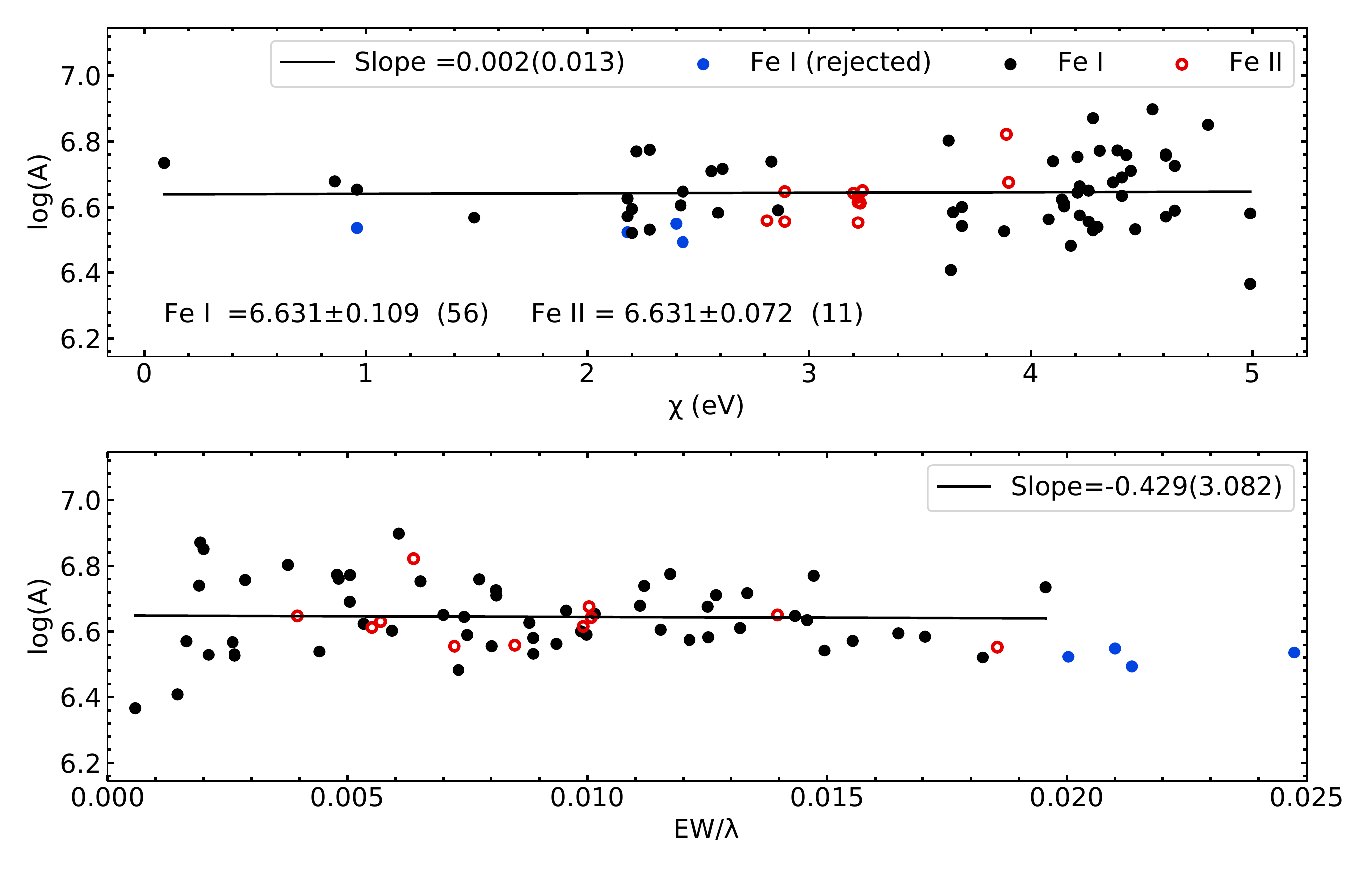}
    \caption{Demonstrates the Fe-Ionization-Excitation equilibrium technique for star 7064.3. The upper panel shows the log of the Fe abundance as a function of excitation potential for Fe I lines (black filled circles) and Fe II lines (red open circles). The lower panel shows the log of the Fe abundances as a function of the reduced equivalent width (EW/$\lambda$). For each panel, the blue filled circles show lines that have a reduced equivalent > 0.02 and are consequently rejected from the analysis. The black lines show the best fit lines to the Fe I abundances for each panel. The text in the upper panel shows the abundance determined using Fe I and Fe II lines, the standard deviation in those abundances and in parentheses, the number of lines used to calculate them.} 
    \label{fig:fe_exc_ion}
\end{figure}

\par In short, BACCHUS derives the effective temperature ($\rm{T_{eff}}$), surface gravity (log $g$), iron abundance ([Fe/H]), and microturbulent velocity ($\rm{v_{micro}}$) using the standard Fe-Ionization-Excitation equilibrium technique (see Figure \ref{fig:fe_exc_ion}) under the assumption of local thermodynamic equilibrium (LTE). Abundances are determined from a $\chi^2$ minimization to synthesized spectra.
\par To determine the stellar parameters, BACCHUS first determines the convolution (accounts for instrumental and rotational broadening) and \vmicro, while fixing the \teff\ and \logg\ to the initial guesses. The \vmicro\ and convolution are solved when the Fe abundances derived from the core line intensity and the equivalent width are in agreement for each line. This ensures there is no correlation between the Fe abundance and the reduced equivalent width (EW/$\lambda$, defined as the equivalent width divided by the wavelength of the line). The equivalent width is calculated by taking the integral of the synthesized spectrum over an automatically selected window, as in \citet{Masseron2016}. Next, BACCHUS fixes the \vmicro\ and convolution and solves for the \Teff\ and \logg. The $\rm{T_{eff}}$ is solved when there is no correlation between the Fe abundance and the excitation potential of the lines and log $g$ is solved when there is no significant offset between the neutral Fe (Fe I) and singly ionized Fe (Fe II) abundances. Here, the Fe abundance is calculated by a $\chi^2$ minimization between the observed spectrum and the synthesized spectrum. Last, BACCHUS uses the previous results to create a grid of 27 model atmospheres and interpolating between them to find the solution where all the criteria for the Fe-Excitation-Ionization equilibrium are met. Each of these steps use up to 80 Fe I lines and 15 Fe II lines. We refer the reader to Section 2.2 of \citet{Hawkins2015} for more information about BACCHUS. 

The error in the \Teff\ is roughly related to the error in the slope of the best fit line for the excitation potential versus Fe abundance plot. The error in \logg\ is roughly related to the error in the Fe I and Fe II abundances. The error in [Fe/H] is the standard deviation in the Fe I abundances. Finally, the error in \vmicro\ is related to the error in the slope of the best fit line for the Fe abundance versus reduced equivalent width plot. We refer the reader to \citet{Masseron2016} for more information on BACCHUS error analysis.

\par We attempted spectral analysis for 33 stars. The stellar parameters were successfully derived for 26 stars in our sample. BACCHUS failed to derive precise stellar parameters for a total of 7 stars. Four of these stars have low SNR ( < 30 $\rm{pixel^{-1}}$). We flag any star with SNR < 30 $\rm{pixel^{-1}}$ that BACCHUS successfully derives parameters. Another star whose calculated SNR is 34 $\rm{pixel^{-1}}$ also failed. Upon further visual inspection of its spectrum, we find it has regions where it is much noisier which causes a large dispersion in the derived Fe II abundances. For the last two stars, BACCHUS was only able to find a solution when it fixes the microturbulence to a set value. However, we do not report these abundances because of the large errors in the derived parameters. In summary, we observed a total of 474 spectra of 40 stars. Seven of these 40 stars are removed during the data reduction process (see Section \ref{sec:uves}). Another seven stars are removed during the spectral analysis for the reasons stated above. This leaves a total of 26 stars for which we derived abundances.

\par The abundances for each element, X, and absorption feature are determined by using the derived stellar parameters to create synthetic spectra with different values of [X/H]. A $\chi^2$ minimization is then performed between the observed spectrum and the synthesized spectra to determine the abundance. BACCHUS automatically rejects lines that are strongly blended. Further, we visually inspect the lines to ensure quality fits to the synthesized spectra. The final elemental abundance is determined by taking the median of the elemental abundances for individual accepted lines. The elemental abundances are scaled relative to solar values from \citet{Asplund2005}. This process was completed for 22 different elements in addition to Fe. 

\begin{table}

\caption{Typical Sensitivities of the [X/H] Abundances on Stellar Parameters}
\label{tab:err}
\begin{tabular}{ccccc}
\hline\hline
[X/H] & $\rm{\sigma_{T_{eff}}}$ & $\rm{\sigma_{log(g)}}$ & $\rm{\sigma_{[Fe/H]}}$ & $\rm{\sigma_{v_{micro}}}$ \\
& ($\pm$ 127 K) & ($\pm$ 0.46) & ($\pm$ 0.16) & ($\pm$ 0.12 \kms\ ) \\
\hline
Fe & $\pm$0.10 & $\pm$0.09 & $\pm$0.03 & $\mp$0.05 \\
Mg & $\mp$0.05 &$\mp$ 0.08 & $\pm$0.02 & $\pm$0.05 \\
Si & $\mp$0.14 & $\pm$0.10 & $\mp$0.02 &$\pm$ 0.04 \\
Ca & $\pm$0.11 & $\mp$0.12 &$\mp$ 0.09 & $\mp$0.07 \\
Ti & $\pm$0.09 & $\mp$ 0.04 & $\mp$0.04 & $\pm$0.03 \\
Mn & $\pm$0.13 & $\mp$0.06 & $\mp$0.09 &$\pm$ 0.04 \\
Co & $\pm$0.04 &$\mp$ 0.02 & $\pm$0.01 & $\pm$0.05 \\
O & $\mp$0.09 & $\pm$0.16 & $\pm$0.08 & $\pm$0.05 \\
Cr & $\pm$0.11 & $\mp$0.07 & $\mp$0.05 & $\mp$0.04 \\
Cu &$\pm$ 0.12 & $\mp$0.10 & $\pm$0.10 & $\pm$0.04 \\
La &$\mp$ 0.05 & $\pm$0.11 &$\pm$ 0.06 & $\pm$0.05 \\
Al & $\mp$0.26 &$\mp$ 0.09 &$\pm$ 0.29 & $\pm$0.05 \\
Na & $\mp$0.01 & $\mp$0.09 & $\pm$0.02 & $\pm$0.05 \\
Ni & $\pm$0.04 & $\pm$0.03 & $\mp$0.00 &$\pm$ 0.01 \\
Ba & $\pm$0.13 & $\pm$0.12 & $\mp$0.01 & $\mp$0.14 \\
Sr & $\pm$0.37 & $\mp$0.28 & $\pm$0.28 & $\pm$0.11 \\
Y & $\mp$0.08 & $\pm$0.12 & $\pm$0.04 & $\pm$0.01 \\
Eu & $\mp$0.11 & $\pm$0.25 &$\pm$ 0.10 & $\pm$0.05 \\
V & $\pm$0.12 & $\mp$0.08 & $\mp$0.03 & $\pm$0.05 \\
Sc & $\mp$0.10 & $\pm$0.11 & $\pm$0.06 & $\pm$0.02 \\
Zr & $\pm$0.12 & $\mp$0.09 & $\pm$0.08 & $\pm$0.05 \\
Nd & $\mp$0.04 & $\pm$0.13 &$\pm$ 0.04 & $\pm$0.05 \\
\hline
\end{tabular}
\flushleft{The typical sensitivities are calculated by measuring the change in abundance [X/H] when the stellar parameters are adjusted by the average error. Columns 2, 3, 4, and 5 shows the change in [X/H] for a change in \teff\ of $\pm$127 K, in \logg\ of $\pm$0.46 dex, in metallicity of $\pm$0.16 dex and in \vmicro\ of $\pm$0.12 \kms, respectively. This is completed for one star (5126.3) in the median of our parameter space (\teff\ = 4785 K, \logg\ = 2.27 dex, [Fe/H] = -0.86 dex, and \vmicro = 1.22 \kms\ ), and has measured abundances in all of our 22 elements besides Li.    }
\end{table}

The uncertainties in the elemental abundances are derived by adding the typical sensitivities of the abundance and the internal error in quadrature. The typical sensitivities for each stellar parameter are found by calculating the change in the abundance after adjusting the parameter by the average error in that parameter. The average error in the \teff, \logg, [Fe/H] and \vmicro\ of our sample is 127 K, 0.46 dex, 0.16 dex and 0.12 \kms, respectively. This process is completed for one star (5126.2) whose parameters are in the center of our parameter space (\teff\ = 4785 K, \logg\ = 2.27 dex, [Fe/H] = -0.86 dex, and \vmicro = 1.22 \kms\ ). The internal error is the line-by-line dispersion in the abundance divide by the square root of the number of lines used. If only one line is used, an internal  error of 0.1 dex is assigned, as in \citet{Hawkins2018b,Howes2016}.

\begin{figure}
    \centering
    \includegraphics[width=\columnwidth]{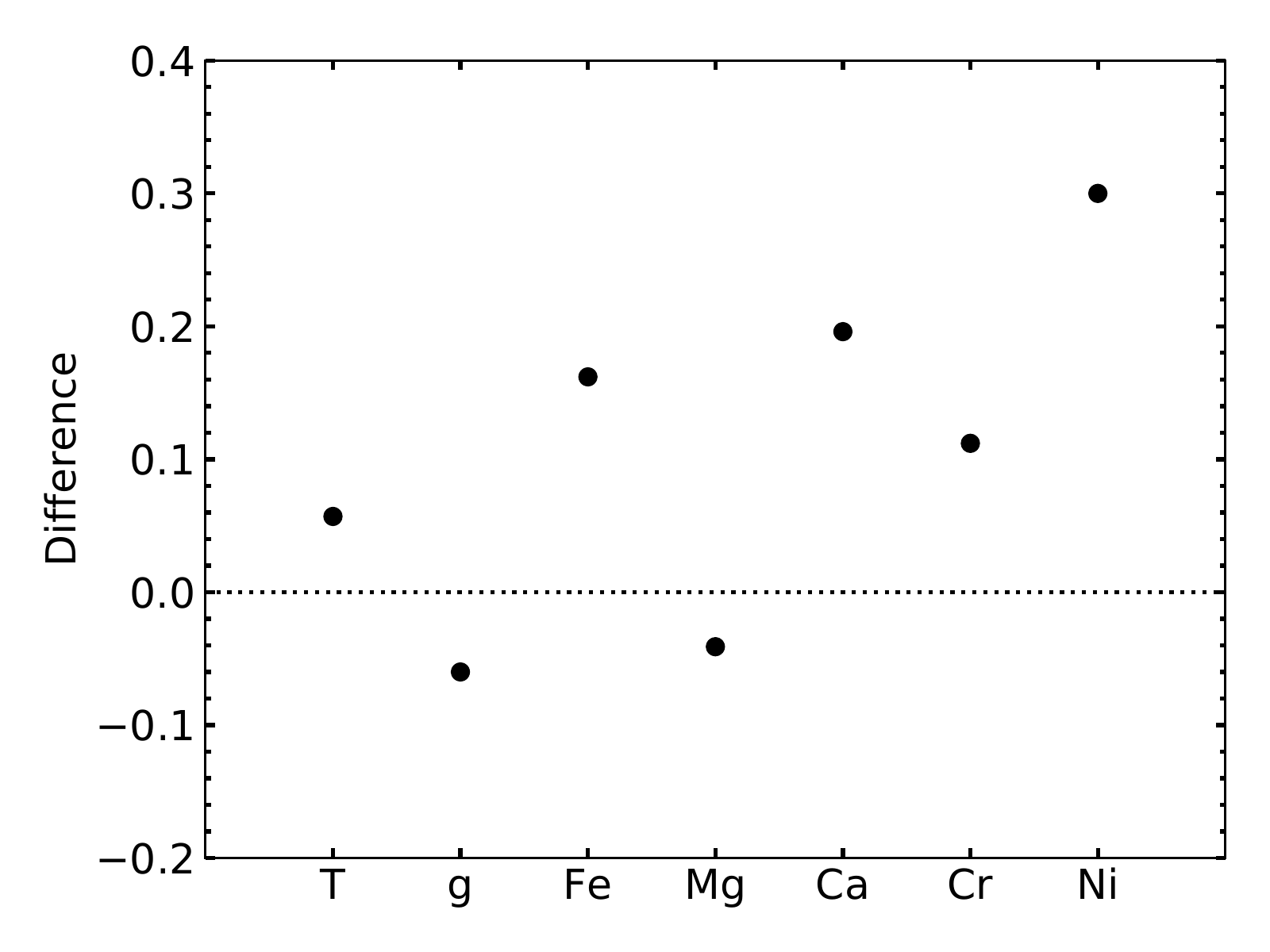}
    \caption{The differences in the derived parameters and elemental abundances between our analysis and the analysis in \citet{Yong2013} for star HE 1506-0113. For each element, the difference reported is $[X/H]_{Yong}-[X/H]_{BACCHUS}$. The two leftmost values show the $\rm{T_{eff}}$ (T) and log $g$. The value for $\rm{T_{eff}}$ is scaled down by a factor of 1000. The elemental abundance differences are used to scale the reported abundances in \citet{Yong2013} to our results. }  
    \label{fig:yong_difs}
\end{figure}

\begin{figure}
    \centering
    \includegraphics[width=\columnwidth]{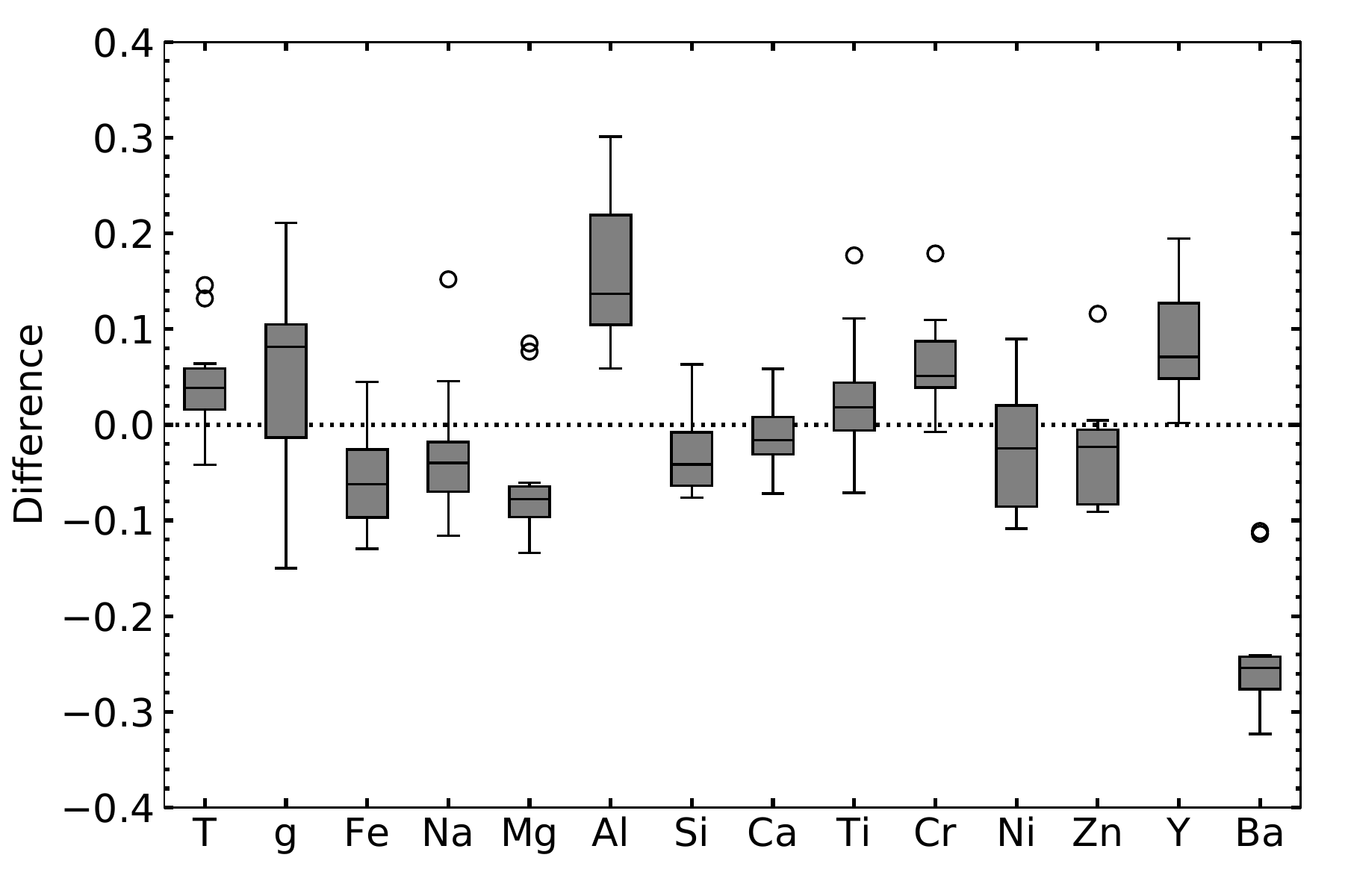}
    \caption{The differences in the derived parameters and elemental abundances between our analysis and the analysis in \citet{Bensby2014} for 12 metal-poor stars observed by VLT/UVES. For each element, the difference reported is [X/H]$_{\mathrm{Bensby}}-$[X/H]$_{\mathrm{BACHHUS}}$. The two leftmost values show the $\rm{T_{eff}}$ (T) and log $g$. The value for $\rm{T_{eff}}$ is scaled down by a factor of 1000. The elemental abundance differences are used to scale our elemental abundances to the results in \citet{Bensby2014}. }  
    \label{fig:Bensby_difs}
\end{figure}
\par The differences in atomic data, adopted solar abundances, and analysis methods cause systematic offsets in abundances between surveys. In order to determine the impact of systematic offsets and accurately compare abundances, a comparative analysis with metal-poor stars from \citet{Bensby2014} and \citet{Yong2013} is performed. We take reduced spectra from the ESO Archive for stars in these samples that are observed with VLT/UVES and have [Fe/H] < -0.5 dex. They are analyzed in BACCHUS and the stellar parameters and elemental abundances are derived as described above. The difference between the reported results in  \citet{Yong2013} and \citet{Bensby2013}, and our analysis are calculated and shown in Figures \ref{fig:yong_difs} and \ref{fig:Bensby_difs}.  The median of the differences for the 12 stars from \citet{Bensby2014} are applied to shift our results to the same scale as the \citet{Bensby2014} results in the figures. The average shift across all elements is -0.05 dex.  The  elemental abundances from \citet{Yong2013} are then shifted according to the differences in Figure \ref{fig:yong_difs} to match our new scale. The average shift across all elements for the results in \citet{Yong2013} is -0.15 dex.

\section{Results} \label{sec:results}

\subsection{Metallicity and Stellar Parameters}

\begin{figure*}
    \centering
    \includegraphics[width=\textwidth]{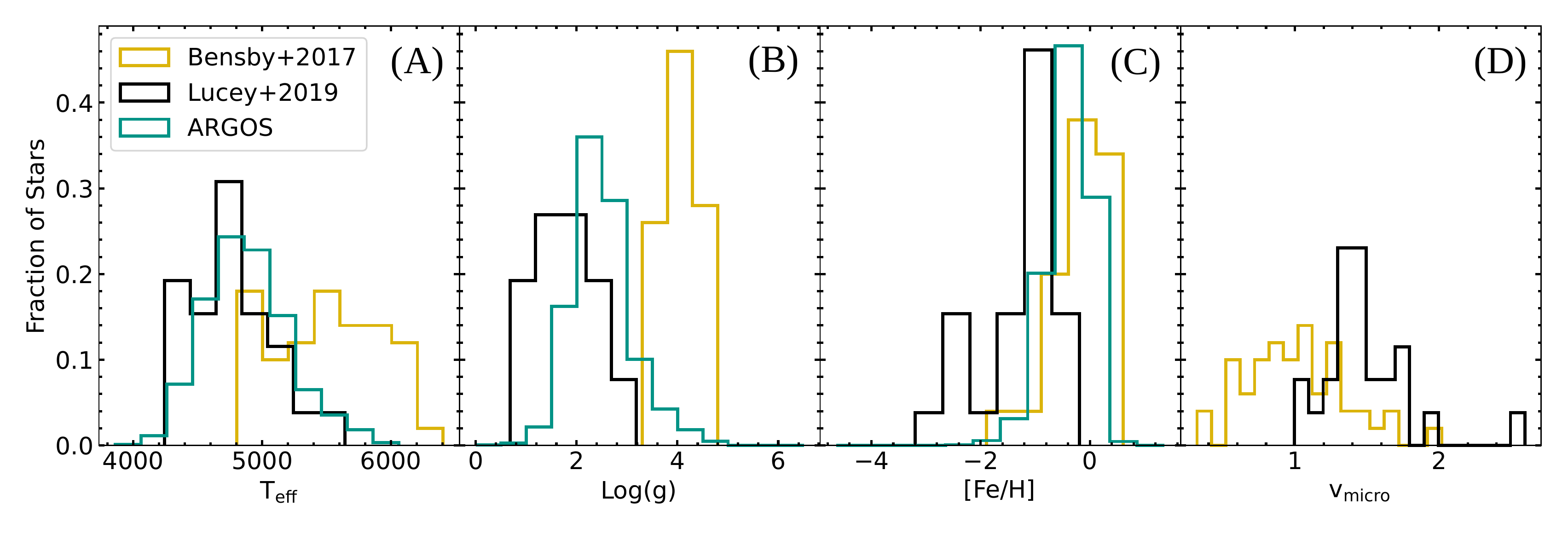}
    \caption{The distribution of stellar parameters for our sample of bulge giant stars compared to the micro-lensed bulge dwarf and subgiant stars from \citet{Bensby2017} and the bulge giant stars in the ARGOS survey \citep{Freeman2013}. Our sample is shown as a black line while the \citet{Bensby2017} sample is shown in gold and the ARGOS survey is shown in green. Our sample focuses on the low-end of the metallicity distribution of the bulge.} 
    \label{fig:params}
\end{figure*}

\par The results of the derived stellar parameters are shown in Figure \ref{fig:params}. The average uncertainties are 127 K, 0.46 dex, 0.16 dex and 0.12 \kms\ for $\rm{T_{eff}}$, log \textit{g}, [Fe/H] and $\rm{v_{micro}}$, respectively. The results shown in Panel C of Figure \ref{fig:params} confirm our SkyMapper selection method has been sufficient to isolate a sample of metal-poor stars in the inner region. We note here that we have derived parameters that are not consistent with the parameters found in the ARGOS survey \citep{Ness2016} for two of our stars (12931.0 and 42011.0). In total, we have four star in common with ARGOS. The average and standard deviation in the differences in the ARGOS parameters and our results are as follows: $\rm{\Delta T_{eff}} =527 \pm 238$K, $\rm{\Delta log(g)}=1.23 \pm 0.76$ dex, and $\rm{\Delta [Fe/H]} = 0.28 \pm 0.33$ dex. We conclude that our parameters are more accurate given the much higher resolution of our spectra.
\par The MDF of the bulge has a mean metallicity of around [Fe/H] = 0 dex. In ARGOS, only about 5\% of stars have metallicities < -1.0 dex \citep{Ness2016}.
\par The average metallicity of our sample is -1.29 dex with a dispersion of 0.74 dex. We find five stars with [Fe/H] < -2.0 dex. The observed metallicity distribution indicates that our sample contains stars that come from the populations associated with the metal-weak thick disk, and the stellar halo.

\subsection{The $\alpha$ elements}
\begin{figure*}
    \centering
    \includegraphics[width=\textwidth]{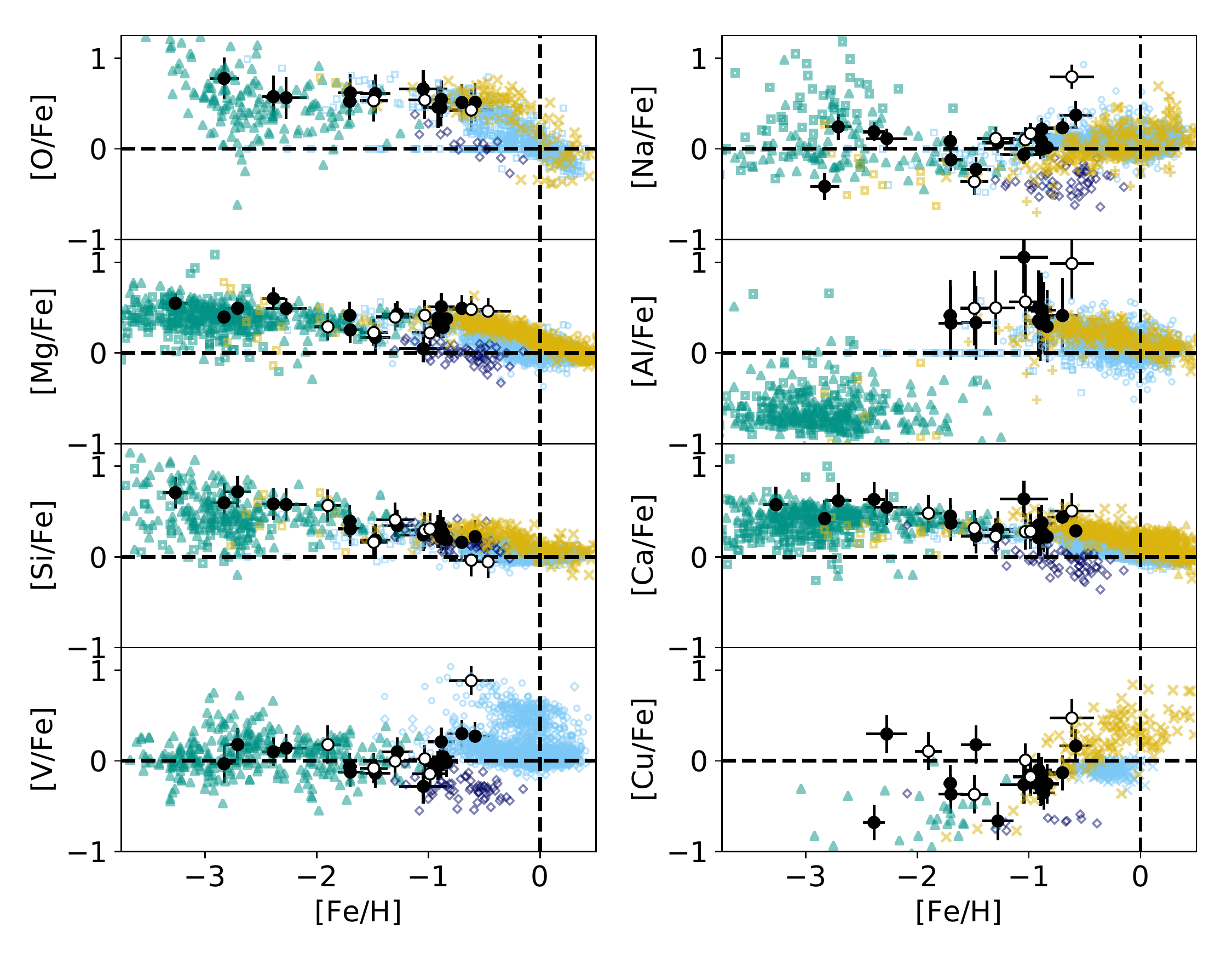}
    \caption{[X/Fe] for the program stars as a function of metallicity for  O, Mg, Si, and V from top to bottom on the left and Na, Al, Ca, and Cu from top to bottom on the right. These elements are $\alpha$ and odd-Z elements. Also shown are abundances for the halo (green), the Large Magellanic Cloud (LMC; dark blue), the disk (light blue) and the bulge (yellow). The program stars with SNR $\geq$ 30 are shown as black filled circles while program stars with SNR between 10 and 30 are shown as black open circles. The halo abundances are from \citet[][green triangles]{Roederer2014}  and \citet[][green open squares]{Yong2013}. The LMC abundances are from \citet[][dark blue open diamonds]{VanderSwaelmen2013}. The disk abundances are from \citet[][light blue open squares]{Bensby2014}, \citet[][light blue open circles]{Adibekyan2012}, and \citet[][light blue open diamonds]{Battistini2015}. The bulge abundances are from \citet[][yellow open squares]{Howes2016}, \citet[][yellow triangles]{Gonzalez2015}, \citet[][yellow xs]{Johnson2014} and \citet[][yellow open circles]{Bensby2017}. }
    \label{fig:alpha}
\end{figure*}

\par Elements that are formed by successive addition of helium nuclei ($\alpha$-particles) are called $\alpha$ elements \citep{Burbidge1957}. $\alpha$ elements are further divided into two categories: hydrostatic and explosive. Hydrostatic $\alpha$ elements (oxygen and magnesium) are primarily formed during the hydrostatic phase of massive stars while the explosive $\alpha$ elements (silicon, calcium, and titanium) are primarily formed during explosive nucleosynthesis of core-collapse supernovae (SNII) \citep{Woosley1995,Woosley2002}.

\subsubsection{O and Mg}

\par The hydrostatic $\alpha$ elements (O and Mg) are formed during the hydrostatic burning phase of massive stars and are dispersed through the ISM through SNII events. Therefore, their ratios relative to Fe are sensitive to the Type Ia time-delay scenario and star formation history.
\par Results in the literature show there is a slight Mg enhancement in the bulge at higher metallicities consistent with a shorter star formation timescale. For example, \citet{Bensby2014} found [Mg/Fe] abundances of micro-lensed dwarf stars in the bulge are slightly higher than in the disk as shown in Figure \ref{fig:alpha}.  \citet{Gonzalez2015} measured similar results in giant bulge stars. Additionally, \citet{Howes2016} found the [Mg/Fe] abundances in the bulge at low metallicity have a large dispersion. 
\par  All of our target bulge stars show enhanced values of [Mg/Fe] relative to the Sun, consistent with previous studies \citep[e.g.,][]{Gonzalez2015,Bensby2017}. At the higher metallicity end, our observed stars show enhancement in [Mg/Fe] relative to the disk, indicating that the bulge may have a shorter star formation timescale. It is interesting to note the low-dispersion of [Mg/Fe] abundances at the low-metallicity end for our observed stars. This is contrary to the dispersion observed in \citet{Howes2016}. We use \citet{Yong2013} to compare our sample's dispersion to the dispersion of the halo. We choose to use \citet{Yong2013} instead of \citet{Roederer2014} because the error analysis in \citet{Yong2013} is very similar to our analysis while \citet{Roederer2014} uses a different method. Therefore, we can more accurately compare the dispersion and precision of our elemental abundances with \citet{Yong2013}. In order to take the difference in sample size into account, we randomly selected stars from \citet{Yong2013} 1000 times and calculated the mean dispersion. For each element, we would select the same number of stars as we had abundance measurements of stars with [Fe/H] < -1.5 dex. We measure a dispersion in [Mg/Fe] at metallicities below -1.5 dex of 0.11 dex and an average error in [Mg/Fe] of 0.13 dex for our sample. For \citet{Yong2013}, we measure an average dispersion of 0.23 dex and an average error of 0.12 dex. Therefore, we do not measure a significant dispersion, while \citet{Yong2013} finds that the halo has significant dispersion in [Mg/Fe]. It is possible that the low dispersion in various elements observed here could partially be attributed to the lack of carbon-enhanced metal-poor (CEMP) stars. However, we note that \citet{Howes2016} observes a large dispersion (possibly from contamination of their bulge sample with halo stars) despite the lack of CEMP stars. In addition, when we remove the CEMP stars from the stellar halo sample from \citet{Yong2013}, we still find an average dispersion that is significantly larger in the halo compared to the bulge targets studied here. 

\par In order to address the impact on [Mg/Fe] of non-LTE (NLTE) at low metallicities, we obtained line-by-line NLTE corrections for Mg from \citet[][see Table \ref{tab:line}]{Bergemann2015}. When these corrections are applied, the mean abundance trend is still consistent with the literature. Although our scatter in the abundances increases, we cannot accurately compare this NLTE scatter to the observed scatter in the halo because \citet{Yong2013} calculates the abundances in LTE and we do not know how they would behave in NLTE.

\par Oxygen abundances are thought to be impacted by an additional mechanism given that it declines more steeply with metallicity in the Galactic disk compared to the other $\alpha$ elements. The two mechanisms that have been proposed to account for this steepening include: 1. stellar wind mass-loss which leads to a decrease in O yield and 2. a steeper IMF at the top-end with increasing metallicity \citep{McWilliam2016}. \citet{Johnson2014} found evidence for a higher [O/Fe] plateau suggesting a top-heavy IMF in the bulge. They also found [O/Fe] enhancement at higher metallicities compared to the disk indicating a shorter star formation timescale that agrees with the results from the other $\alpha$ elements.
\par As shown in Figure \ref{fig:alpha}, our observed O abundances show an agreement with the disk plateau indicating that the bulge and disk may have the same IMF. Interestingly, we do not measure a significant dispersion in [O/Fe] for stars on our sample with [Fe/H] < -1.5 dex. Our measured dispersion is 0.09 dex, while our average error is 0.22 dex. 

\subsubsection{Si and Ca}

\par The $\alpha$ elements Si and Ca are primarily formed during the explosive nucleosynthesis of SNII events. SNII yield more of these $\alpha$ elements than Fe while SNIa yield more Fe than these elements. Therefore, Si and Ca are sensitive to the SNIa time-delay scenario and star formation history.
\par There is ample evidence for an enhanced [X/Fe] ratio for Si and Ca at roughly solar metallicities in the Galactic bulge, indicating higher rates of star formation than in the disk. For example, \citet{Johnson2014}, \citet{Howes2016} and \citet{Bensby2017} all measured Ca and Si abundances for stars in the bulge and found that they are enhanced relative to the disk. This enhancement starts around metallicities of -1.00 dex. Below those metallicities, \citet{Howes2016} found a large dispersion in [Si/Fe] and [Ca/Fe], and an overall enhancement relative to the Sun.

\par Our derived elemental abundances for Si and Ca are consistent with the literature and our observed hydrostatic $\alpha$ abundances. However, the dispersion at low metallicities is much lower for our program stars than in the halo and in \citet{Howes2016}. The average dispersions of Si in \citet{Yong2013} is more than twice the measured dispersions of our sample. It is possible that the large dispersion in \citet{Howes2016} is due to contamination from halo stars. Though, studying their kinematics will allow us to confirm if that is the case. Figure \ref{fig:rz} shows there are several stars with Galactic radii > $\sim$ 6 kpc, according to the \citet{Bailer-Jones2018} distance catalog. We also note that \citet{Fulbright2007} found similar results and measured a ``starkly" small scatter in Si, Ca and Ti abundances in the the bulge compared to the halo.

The level of $\alpha$ enhancement observed in these low metallicity stars provides evidence that these stars are not from an accreted dwarf galaxy. The high level of enhancement in Ca for four of our stars with [Fe/H] < -2 dex relative to the Galactic disk plateau at [Fe/H] < -2 dex is consistent with a flat IMF \citep[e.g.,][]{Johnson2013a}. Our Ca abundances are also consistent with \citet{Duong2019} who also measure high levels of Ca enhancement in metal-poor bulge stars. At the higher metallicity end the [$\alpha$/Fe] values have similar trends to those seen in the disk, but are more enhanced likely from a shorter star formation timescale. 

\par Line-by-line NLTE corrections for Si were obtained from \citet{Bergemann2013}. When these corrections are applied to our Si abundances, the mean abundance trend is still consistent with the literature. Although our scatter in the abundances increases, we cannot accurately compare this NLTE scatter to the observed scatter in the halo because \citet{Yong2013} calculates the abundances in LTE.

\subsection{The odd-Z elements}
Odd-Z elements are any element with an odd atomic number (and therefore could not have been produced by successive addition of $\alpha$-particles). Of the odd-Z elements, we measured elemental abundances for sodium (Na), aluminum (Al), vanadium (V), copper (Cu) and lithium (Li). 

However, we were only able to measure Li abundances for four of our stars. These stars all have A(Li)\footnote{Li abundances are reported in the standard way, where A(Li) = $\rm{log}\left(\frac{N_{Li}}{N_{H}}\right)$+12, where $N_{Li}$ and $N_H$ are the number of lithium and hydrogen atoms per unit volume, respectively. }between 0.5 and 1 dex. These abundances are roughly what are expected for giant stars with the \Teff\ probed here. The uncertainties in the Li abundances are calculated in the same manner as the other elements. However since the selected median star (5126.3) for the typical sensitivities did not have a measured Li abundance, we calculated the typical sensitivities for Li using another star, 42011.0  (\teff\ = 4559 K, \logg\ = 0.98 dex, [Fe/H] = -2.65 dex, and \vmicro = 1.45 \kms\ ). Analysis of this star resulted in the following typical sensitivities: $\rm{\sigma_{T_{eff}}}$ = 0.043 dex, $\rm{\sigma_{log(g)}}$ = 0.016 dex, $\rm{\sigma_{[Fe/H]}}$ = 0.034 dex, and $\rm{\sigma_{v_{micro}}}$ = 0.051 dex.  The other odd-Z elements are  will be discussed separately below. 

\subsubsection{Na}
\par Results in the literature show that there is no significant difference between the trends in [Na/Fe] as a function of metallicity between the bulge and the disk. The lowest metallicity stars in our sample show enhanced [Na/Fe] relative to the Sun. However, we note the low dispersion and enhancement relative to \citet{Howes2016}. Our sample also shows a lower dispersion at [Fe/H] <-1.5 dex ($\rm{\sigma_{Na}}$ = 0.22 dex) than the halo sample of \citet{Yong2013} ($\rm{\sigma_{Na}}$ = 0.42 dex). This difference in dispersion is significant given the average error of our [Na/Fe] abundances is 0.12 dex and \citet{Yong2013} reports an average error of 0.15 dex.  At the higher metallicity end of our sample the [Na/Fe] abundances are disk-like. As shown in Figure \ref{fig:alpha}, there is one clear outlier in [Na/Fe]. This star is an outlier in other elements as well and will be discussed further in Section \ref{sec:cluster}
\subsubsection{Al}
\par Results in the literature show Al's $\alpha$-like abundances in the bulge at metallicities above $\sim$-1 dex. In this range, the bulge shows a slight enhancement of [Al/Fe] with respect to the disk. This is consistent with a shorter star formation timescale. At the low metallicity end ([Fe/H] < $\sim$-1 dex), \citet[][yellow squares in Figure \ref{fig:alpha}]{Howes2016} found the bulge has [Al/Fe] similar to the halo.  Unfortunately, we are unable to measure Al abundances for our lowest metallicity stars due to low SNR. The [Al/Fe] ratios for our higher metallicity stars are consistent with previous results showing slight enhancement compared to the disk. There are two clear outliers with [Al/Fe] $\sim$ 1 dex. Large Al enhancement paired with low [Mg/Fe] is a signature of second generation globular cluster stars \citep{Gratton2001,Carretta2004,Ramirez2002,Ramirez2003,Lind2015}. Therefore, we explore the possibility of globular cluster origin for these stars in Section \ref{sec:cluster}. 
\subsubsection{V}

\par Unfortunately, there has been very little work measuring V abundances in the Galactic bulge. However, we can still compare our results to literature values in the disk, halo and the LMC. As shown in Figure \ref{fig:alpha}, [V/Fe] is roughly flat with metallicity in the halo, while showing an $\alpha$-like slope in the disk and LMC. Overall, the [V/Fe] abundances of our target stars are consistent with those seen in the halo. 

\subsubsection{Cu}
\par Consistent with a shorter star formation timescale, \citet{Johnson2014} measured [Cu/Fe] enhancement in the bulge relative to the disk at metallicities above $\sim$ -1 dex. Our results support this conclusion. At lower metallicities ([Fe/H] < -1 dex) our bulge stars have a large dispersion in [Cu/Fe]. However, further measurements of Cu abundances for stars throughout the Milky Way will be needed to constrain its production site. 
\subsection{The Fe-peak elements}
\begin{figure*}
    \centering
    \includegraphics[width=\textwidth]{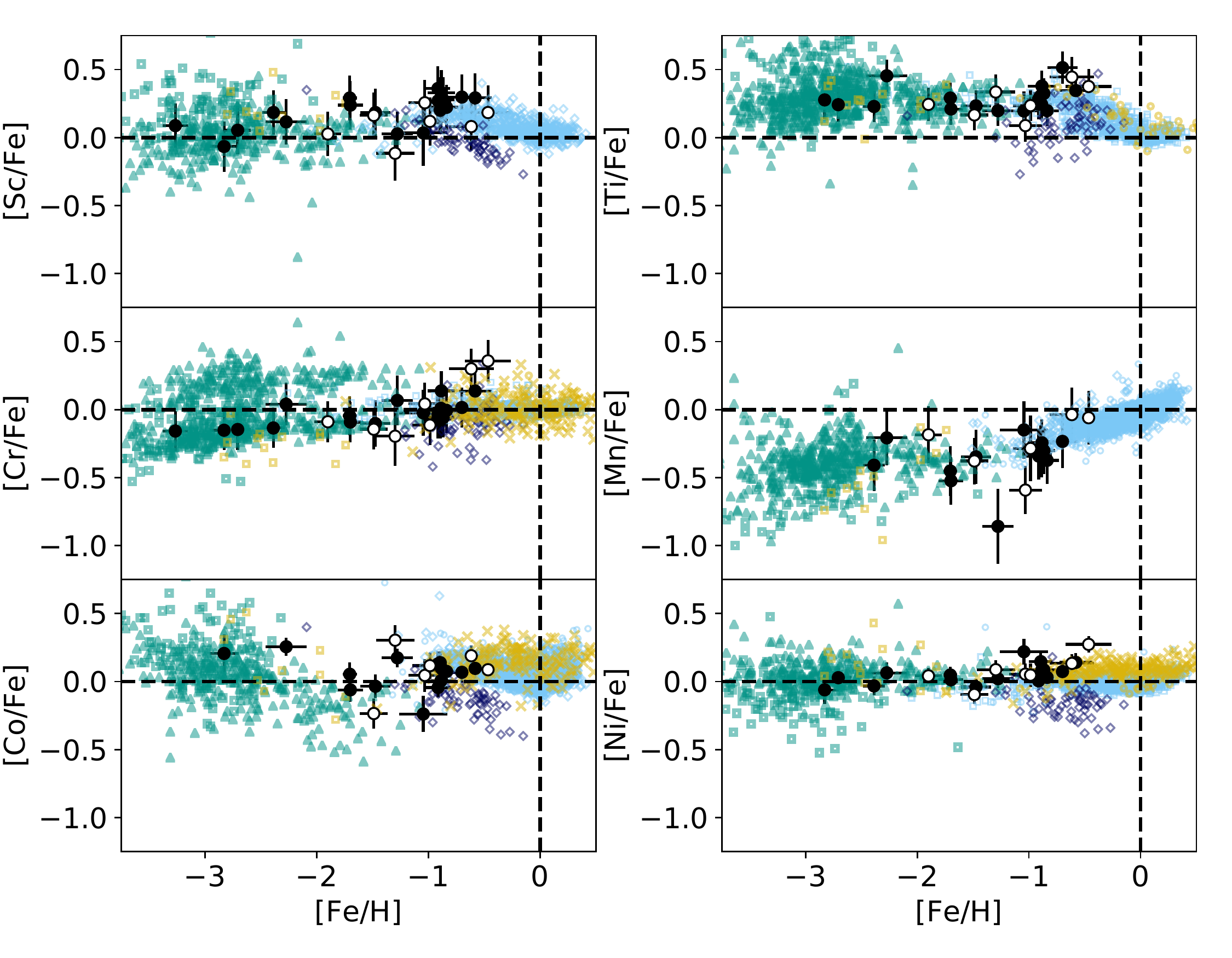}
    \caption{[X/Fe] for the program stars as a function of metallicity for  Sc, Cr, and Co from top to bottom on the left and Ti, Mn, and Ni from top to bottom on the right. These elements are Fe-peak elements. The symbols are the same as in Figure \ref{fig:alpha} with the addition of copper abundances for the disk from \citet[][light blue xs]{Reddy2003}. }
    \label{fig:fe}
\end{figure*}

\par The Fe-peak elements (chromium, nickel, scandium, titanium, manganese, zinc and cobalt) are formed in a variety of ways \citep{Iwamoto1999,Kobayashi2006,Nomoto2013} but are largely dispersed into the ISM in ways similar to iron. Therefore, the Fe-peak elements generally trace iron with small variations except for Mn. We discuss each of the Fe-peak elements shown in Figure \ref{fig:fe} separately.

\subsubsection{Sc}
There has not been much previous work measuring scandium (Sc) abundances in the bulge. \citet{Howes2016} found low metallicity stars in the bulge have [Sc/Fe] abundances similar to the halo. Our results support this conclusion. At [Fe/H] < $\sim$ -1.00 dex,  our sample has [Sc/Fe] abundances that are consistent with the halo, and are roughly flat with values around the solar value.  At higher metallicities ([Fe/H] > $\sim$-1 dex), Sc appears to have $\alpha$-like trends in the disk and the LMC. We find that our results have an $\alpha$-like trend and show a slight enhancement compared to the disk. This enhancement is consistent with a shorter star formation timescale in the bulge.

\subsubsection{Ti}
 Titanium (Ti) behaves similarly to the $\alpha$ elements. Its yield from SNII is higher than the yield from SNIa with respect to Fe. Therefore, it is sensitive to the SNIa time-delay scenario and star formation history. Often, it is thought of as an $\alpha$ element, however it is not formed through the successive addition of $\alpha$ particles. Therefore, we do not categorize it as an $\alpha$ element.

 \citet{Bensby2017}  and \citet{Howes2016} successfully measured Ti abundances for bulge stars for a range of metallicities. Their measured [Ti/Fe] abundances are enhanced relative to the disk at [Fe/H] > ~-1 dex. This indicates a shorter star formation timescale which agrees well with the results for $\alpha$ elements in the bulge. 
 
 Our measured [Ti/Fe] abundances are consistent with the literature. For [Fe/H] > ~-1 dex, we measure [Ti/Fe] that is enhanced relative to the disk, indication a shorter star formation timescale in the bulge. For [Fe/H] < -1.5 dex our Ti abundances show a lower dispersion than the halo. The average dispersions of Ti in \citet{Yong2013} is more than twice the measured dispersions of our sample.
 
Line-by-line NLTE corrections for Ti were obtained from \citet{Bergemann2011}. When the Ti corrections are applied, [Ti/Fe] is enhanced to values $\sim$ 0.75 dex. Because our comparative literature samples all have LTE abundances, we are unable to accurately compare our NLTE abundances to known stellar populations. Therefore, we do not draw conclusions about the origins of these stars from their NLTE abundances.

\subsubsection{Cr}
Chromium (Cr) abundances roughly track Fe abundances. Therefore, [Cr/Fe] as a function of metallicity is largely flat and centered at the solar value. As shown in Figure \ref{fig:fe} this is largely what has been seen in the halo and disk populations \citep{Roederer2014,Yong2013,Bensby2014}. We find that our sample has elemental abundances consistent with the halo and disk populations. It is interesting to note that our lowest metallicity stars all have [Cr/Fe] deficient relative to the Sun. We also note the low dispersion of our sample relative to the halo. The dispersion in [Cr/Fe] for our stars with [Fe/H] < -1.5 dex is 0.06 dex and the average error is 0.15 dex while the average dispersion for the \citet{Yong2013} sample is 0.16 dex  and the average error is 0.12 dex. This indicates these stars are likely from a distinct population, or that the halo is more chemically homogeneous towards the Galactic center. Overall, our results are consistent with the literature in the bulge. The combined results from our study and \citet{Howes2016} indicate possible Cr deficiency in the bulge at low metallicities.

\subsubsection{Mn}
\par Mn is thought to be produced in SNII and SNIa events. Theoretical work indicates SNII events under-produce [Mn/Fe] at roughly -0.3 to -0.6 dex regardless of metallicity \citep{Kobayashi2006} while SNIa events produce yields of Mn that increase with metallicity \citep{Kobayashi2015}. These theoretical results indicate that [Mn/Fe] is sensitive to the SNIa time-delay scenario. Opposite to the $\alpha$-like trends, [Mn/Fe] as a function of metallicity has a plateau below solar values and it begins to increase at the ``knee". This trend has been observed in the Galactic halo and disk \citep{Roederer2014,Yong2013,Adibekyan2012}. It is debated whether the observed trend of [Mn/Fe] as a function of metallicity in the Milky Way is astrophysical or due to metal-dependent NLTE effects in stellar atmospheres.
\par There has been little work measuring Mn abundances in the Galactic bulge. Our elemental abundances measured for the lowest metallicity stars in our sample are consistent with the observed LTE halo abundances. At the higher metallicity end of our sample [Mn/Fe] shows slight deficiency relative to the disk consistent with a shorter star formation timescale. Further work on Mn abundances in the bulge are desired.

Mn NLTE corrections are sourced from \citet[][see Table \ref{tab:line}]{Bergemann2008}. When the abundance corrections are applied, we see [Mn/Fe] is flat across our metallicity range. This is the same trend seen in the NLTE Mn results from \citet{Battistini2015}. Because our comparative literature samples all have LTE abundances, we are unable to accurately compare our NLTE abundances to known stellar populations. Therefore, we do not draw conclusions on the origins of these stars from their NLTE abundances.

\subsubsection{Co}
\par NLTE corrections are also determined for Co from \citet[][see Table \ref{tab:line}]{Bergemann2010}. These corrections result in a trend that continues to increase with decreasing metallicity below [Fe/H]  $\sim$ -1 dex. However, because our comparative literature samples all have LTE abundances, we are unable to accurately compare our NLTE abundances to known stellar populations. Therefore, we do not draw conclusions on the origins of these stars from their NLTE Co abundances.

\subsubsection{Ni}
Similar to Cr, nickel (Ni) abundances roughly track Fe abundances and we expect [Ni/Fe] as a function of metallicity to be roughly flat. Interestingly, [Ni/Fe] deficiencies have been measured for $\alpha$-poor systems like the LMC \citep{VanderSwaelmen2013} and slight [Ni/Fe] enhancements have been observed in the bulge at metallicities above $\sim$ -1 dex \citep{Johnson2014,Bensby2017}. Our results provide further evidence for this enhancement. Our low metallicity stars have [Ni/Fe] abundances consistent with the halo.

\subsection{The Neutron-Capture elements}
\begin{figure*}
    \centering
    \includegraphics[width=\textwidth]{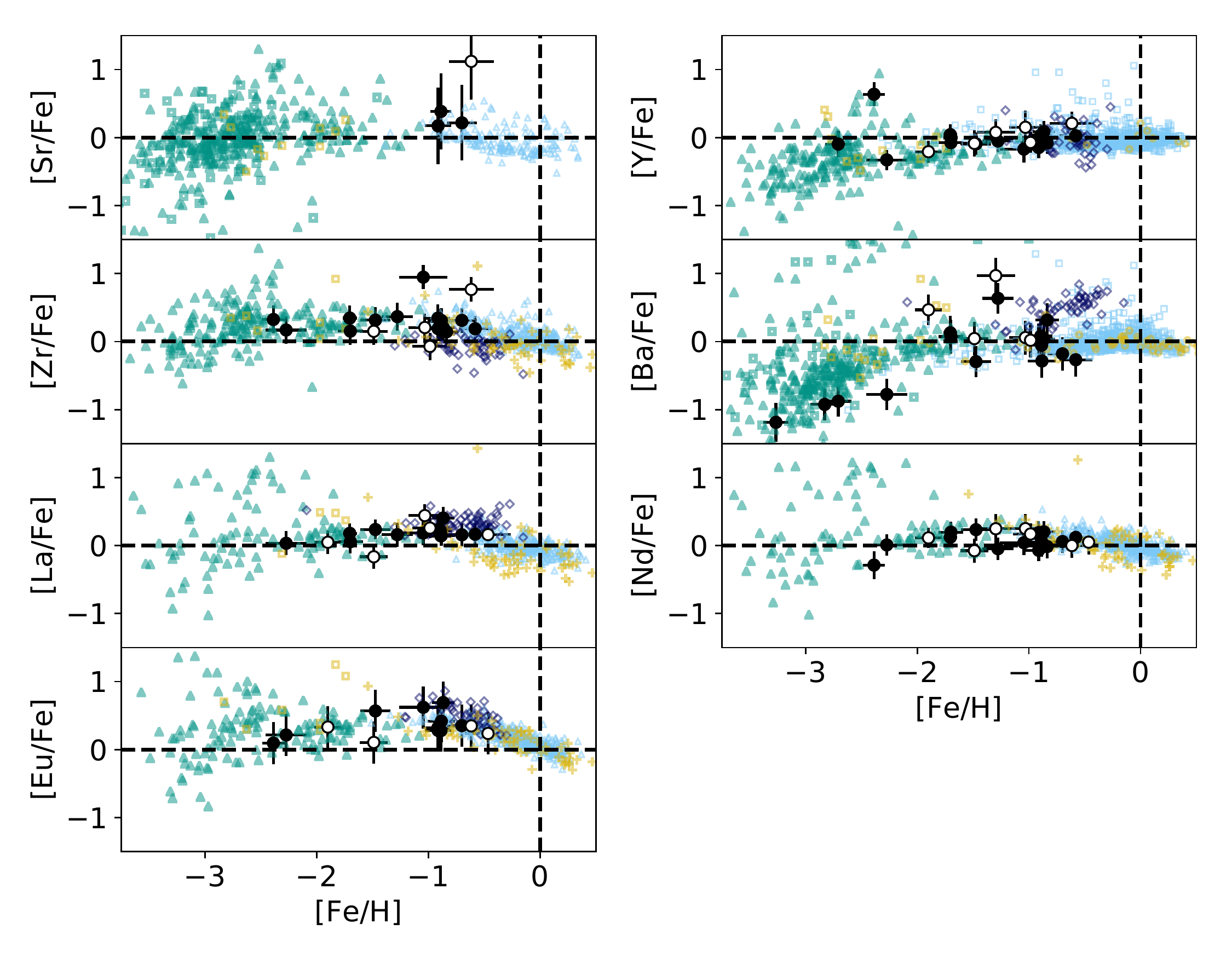}
    \caption{[X/Fe] for the program stars as a function of metallicity for  Sr, Zr, La and Eu from top to bottom on the left and Y, Ba, and Nd from top to bottom on the right. These elements are neutron-capture elements. The symbols are the same as in Figure \ref{fig:alpha} with the addition of disk abundances from \citet[][light blue open triangles]{Battistini2016} and bulge abundances from \citet[][yellow crosses]{Johnson2012}. }
    \label{fig:neutron}
\end{figure*}
 Enhancements of s- and r-process material has frequently been observed in very metal-poor halo stars \citep[e.g.,][]{Sneden2002,Barbuy2009,Masseron2010, Chiappini2011,Sakari2018}. These enhancements are thought to either a result of enrichment from an early generations of spinstars \citep{Pignatari2008} or neutron star mergers \citep{Lattimer1974,Rosswog2014,Lippuner2017}, or from mass-accretion from an AGB binary companion \citep{Abate2015}.  Detections of neutron-capture element enhancements among metal-poor stars in the bulge have been rare, given the rate at which this stars appear in the halo \citep{Koch2019}. Currently, there are only three known metal-poor s- and/or r-process enhanced stars in the bulge \citep{Johnson2013b,Koch2019}.
\subsubsection{r-process elements}
Europium (Eu) is almost purely produced through r-processes \citep{Burbidge1957,Bisterzo2011}. The observed decline of [Eu/Fe] with metallicity observed in the disk (see Figure \ref{fig:neutron}) is thought to be from the Type Ia time-delay scenario given its $\alpha$-like appearance. Similar to $\alpha$ elements, the theorized shorter star formation timescale would lead to an enhancement of Eu in the bulge relative to the disk. However, there is no evidence of this enhancement \citep{McWilliam2016}. At metallicities above $\sim$ -1 dex, our observed [Eu/Fe] as a function of metallicity are consistent with the disk while at lower metallicities, they are consistent with the halo. 
\subsubsection{s-process elements}
\par A significant portion of the production of neodymium (Nd) and zirconium (Zr) are through r-processes even though they are largely thought of as s-process elements. Therefore, the behavior of [X/Fe] of Nd and Zr as a function of metallicity is similar to Eu. Just as with Eu, Nd and Zr are expected to be slightly enhanced in the bulge if there is $\alpha$ enhancement with respect to the disk. Again, there is no evidence for this enhancement in the bulge \citep{McWilliam2016}. Our observed [Nd/Fe] and [Zr/Fe] abundances are consistent with the disk at metallicities above $\sim$ -1 dex and with the halo below. The outlying stars with high levels of Zr enhancement also are enhanced in Al and are discussed further in Section \ref{sec:cluster}.
\par Elements thought to be almost solely created through s-processes (Sr, Y, Ba, and La) show roughly flat ratios of [X/Fe] as a function of metallicity. The slight decrease of [X/Fe] as a function of metallicity shown in the disk observations for Sr and La (see Figure \ref{fig:neutron}) is not well understood \citep[e.g.,][]{Cristallo2011}. Regardless, Figure \ref{fig:neutron} shows that our target stars are consistent with s-process abundances in the halo at low metallicities and the disk at higher metallicities ([Fe/H] >-1 dex). The behavior of [La/Eu] as a function of metallicity for our stars indicates higher levels of r-process material enrichment than s-process material relative to the thin disk. This is consistent with previous results in the bulge \citep{McWilliam2010b,Johnson2012}.

\section{Discussion and Summary} 
\label{sec:cluster}
The MDF of the bulge indicates multiple populations \citep{Ness2013,Zoccali2017,Bensby2017}. The most metal-poor population of the bulge has recently become a focus of interest \citep[e.g.,][]{Howes2016,Howes2014,Howes2015,Schlaufman2014,GarciaPerez2013}. The origin of these stars are under debate. Whether this population is mostly halo interlopers with eccentric orbits, accreted material or some of the oldest stars in the Universe is still yet to be determined. \citet{Howes2015} found that only 7 out of 10 metal-poor bulge stars have orbits confined to the bulge. It may be possible to determine the origin of these stars by studying their chemical composition. In this study, we successfully targeted metal-poor bulge stars using SkyMapper photometry. We obtained high-resolution spectra of 40 targets using VLT/UVES. These spectra were reduced in the standard way. BACCHUS was used to determine the stellar parameters and abundances of 22 elements for 26 stars.

We find our targets to have an average metallicity of -1.29 dex with dispersion 0.74 dex. To discuss the results, we divide our targets into two groups, high metallicity ([Fe/H] > -1.5 dex) and low metallicity ([Fe/H] <-1.5 dex). In general, the high metallicity stars have elemental abundances consistent with other bulge studies at those metallicities \citep[e.g.,][]{Gonzalez2015,Johnson2014,Bensby2017,Johnson2012}. The $\alpha$ abundances for the high metallicity stars are consistent with a high SFR in the bulge relative to the thick disk. We find that two high metallicity stars with unusually high Al abundances which we discuss shortly. In general, we find the elemental abundances of the low-metallicity stars are consistent with halo abundances. The $\alpha$ abundances of these stars are similar to the most $\alpha$-enhanced stars in the halo. This $\alpha$ enhancement indicates that these stars are from a more massive system than a typical dwarf spheroidal galaxy and therefore, not likely to be from an accreted dwarf galaxy. We find four stars with [Ca/Fe] enhancement higher than the Galactic disk plateau. This indicates these stars were enhanced from a population with a more top-heavy IMF \citep[e.g.,][]{Johnson2013a}. We also find the the dispersion in the $\alpha$ and odd-Z elements is generally lower than the average dispersion in the halo populations from \citet{Yong2013}. This indicates the metal-poor bulge is a distinct bulge population, or that the halo is more chemically homogeneous closer to the center of the Galaxy. However, our sample size is relatively small, with only 26 stars, and we caution drawing conclusions from the observed low dispersion.

\begin{figure}
    \centering
    \includegraphics[width=\columnwidth]{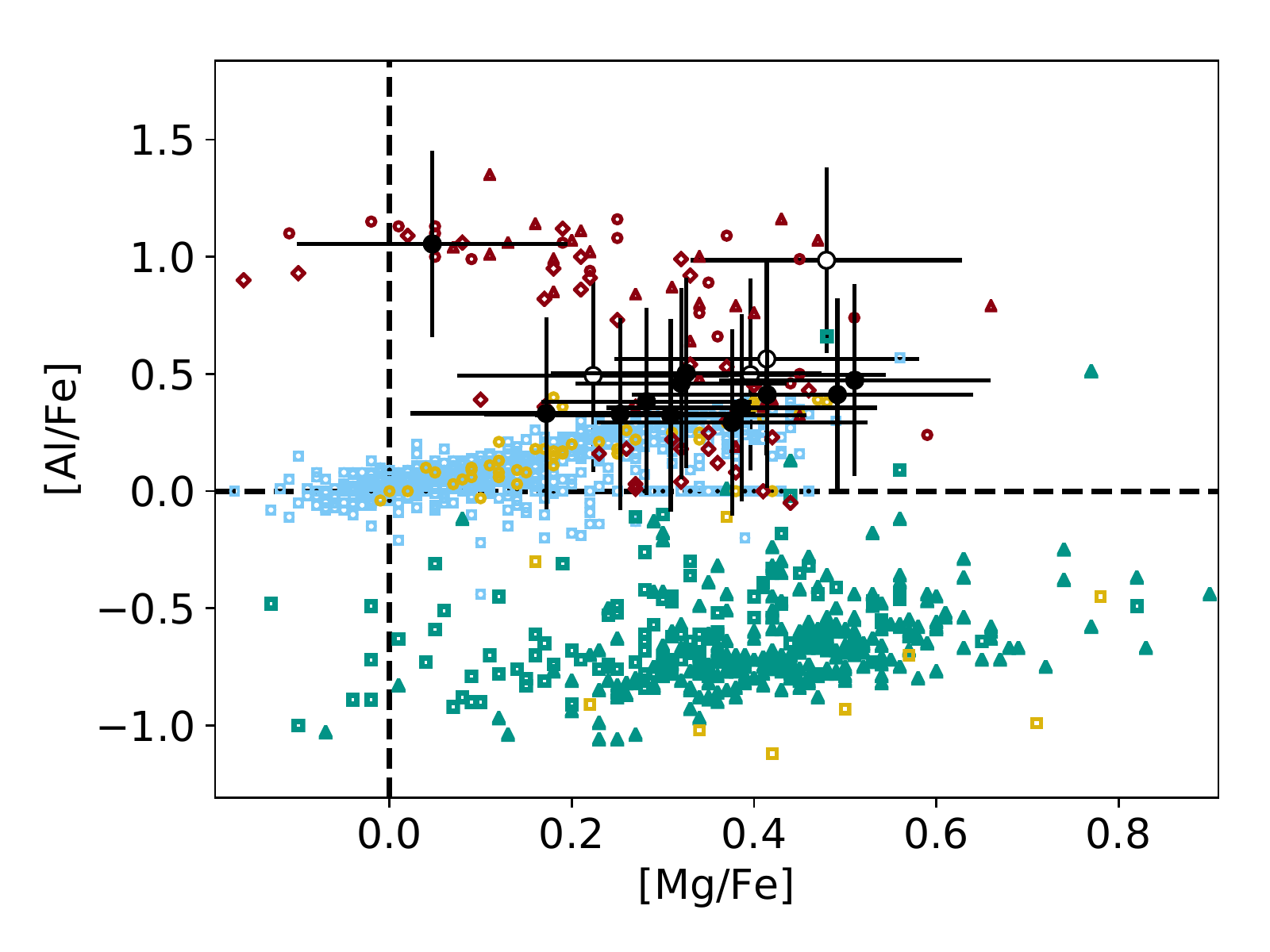}
    \caption{The [Al/Fe] abundances as a function of [Mg/Fe] for our target stars. The symbols are the same as in Figure \ref{fig:alpha} with the addition of elemental abundances for star in the globular clusters NGC4833 (red circles), NGC7089 (M2; red triangles) and NGC2808 (red diamonds) from \citet{Pancino2017}.}
    \label{fig:al_mg}
\end{figure}

\begin{figure}
    \centering
    \includegraphics[width=\columnwidth]{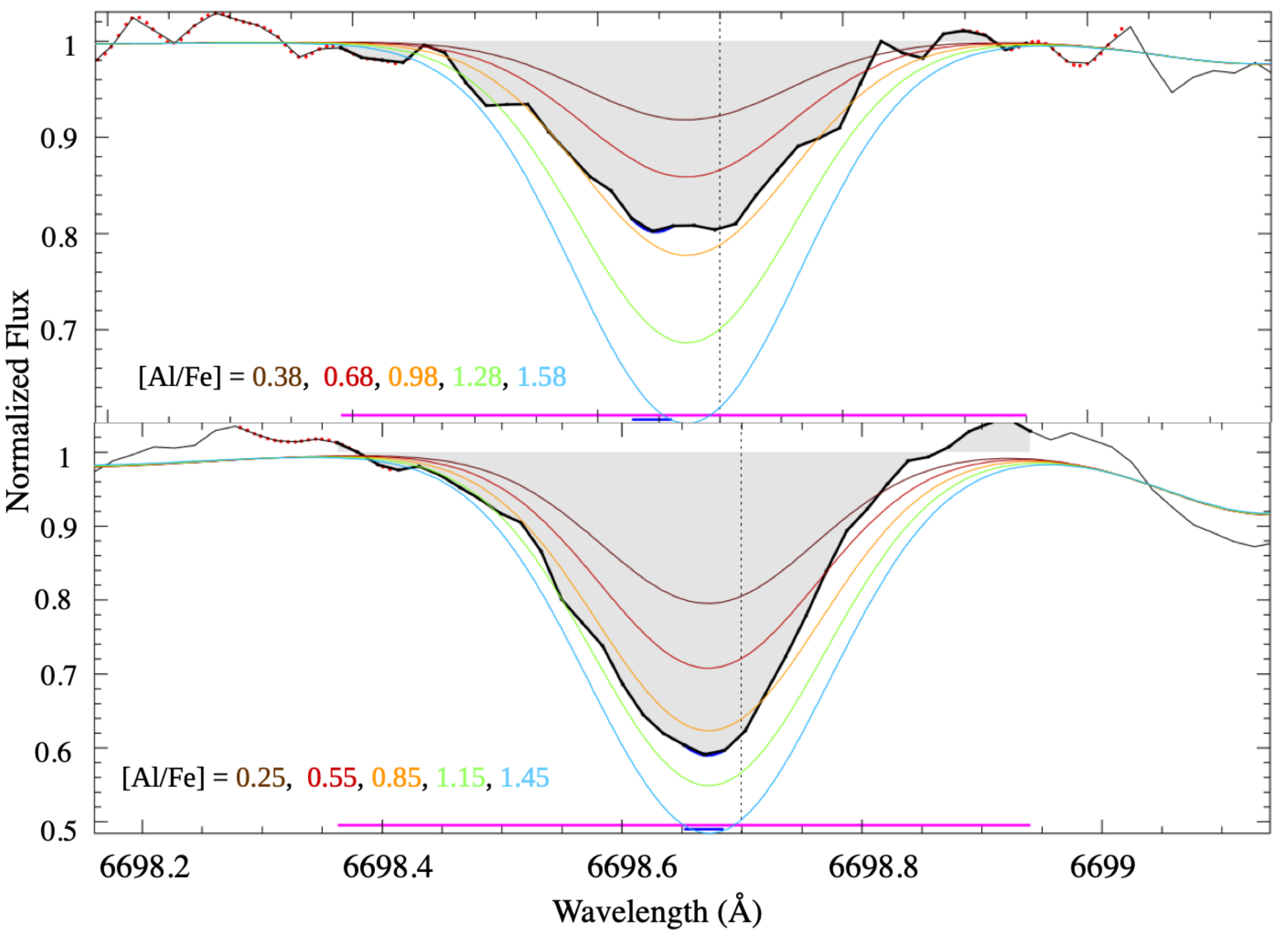}
    \caption{The Al line at 6698.67~\AA\ in the spectra of 6531.3 (top) and 6805.0 (bottom) along with synthesized spectra with varying Al abundances. [Al/Fe] for each synthesized spectrum is given in the bottom right, in order of increasing [Al/Fe]. These lines clearly show these stars have enhanced [Al/Fe] $\sim$ 1 dex, consistent with Figure \ref{fig:al_mg}.}
    \label{fig:al}
\end{figure}

We also find two stars (6805.0 and 6531.3) that have unusually high Al abundances at around the 2$\sigma$ level. It is thought that the bulge may be partially built from dissipated globular clusters \citep{Kruijssen2015,Shapiro2010,Bournaud2016}. \citet{Schiavon2017} found a population of stars in the Galactic bulge whose chemistry is consistent with the known chemical signature of globular cluster stars. These stars are nitrogen rich. \citet{Fernandez2017} found 5 stars with chemistry consistent with globular cluster stars and have highly eccentric orbits that pass through the bulge   Some second generation (SG) globular clusters (GC) stars are thought to have a unique chemical signature in that they show a Mg-Al anti-correlation. That is to say, they are more enhanced in Al than expected given their Mg abundance. Figure \ref{fig:al_mg} demonstrates that the two stars have Mg and Al abundances consistent with SG GCs\footnote{We note here that the reported abundances for Mg and Al in the online table for \citet{Pancino2017} are mislabeled. The column labeled Mg is actually the Al abundances from the Gaia-ESO survey \citep{Gilmore2012,Randich2013} fourth data release and vice versa. }. We note that the large errors bars are a result of the strong sensitivity to the stellar parameters (see Table \ref{tab:err}) and the strength of the Al lines clearly show high levels of Al enhancement (see Figure \ref{fig:al}). It is also known that GCs frequently show a Na-O anti-correlation \citep{Gratton2004}. 6805.0 has Na and O abundances consistent with those seen in the globular cluster $\omega$ Centauri which has stars with -2.0 dex < [Fe/H] < -0.7 dex \citep{Marino2011}. 6531.3 does not show Na enhancement, so we suggest caution with assuming this star is a SG GC star. The chemistry of these two stars indicate that part of the metal-poor population of the bulge could be made up of dissipated globular clusters. There are no known globular clusters within an angular separation of one degree of these stars.

In conclusion, we find evidence that the metal-poor stars in the bulge are not halo interlopers or accreted from a dwarf galaxy. Although these stars have elemental abundances consistent with those seen in the halo, the dispersion of the elemental abundances are not. We also find evidence that a portion of the population of the metal-poor stars in the bulge may have come from globular clusters. More precise orbits with \gaia\ Data Release 3 and an increase in the sample size could definitively rule out halo or accreted origin for the metal-poor population in the bulge. If these stars are not halo-interlopers or accreted, then they are likely to be some of the oldest known stars in our Galaxy and could be used to study the early universe. For the second part of the COMBS survey, we plan to determine the orbits of these stars along with the 640 GIRAFFE spectra in order to determine if the metal-poor bulge population is confined to the bulge.

\appendix
\section{Online Tables}
Sections of two tables available online are shown in Table \ref{tab:abunds} and \ref{tab:line}. Table \ref{tab:abunds} provides the derived stellar parameters and elemental abundances for each star. Table \ref{tab:line} gives the line-by-line abundance for each star and element, along with the oscillator strength (log($gf$), wavelength, and excitation potential ($\chi$) of each line.

\begin{table*}

\caption{Stellar parameters and elemental abundances for 26 metal-poor bulge stars.}
\label{tab:abunds}
\begin{tabular}{ccccccccccc}
\hline\hline
Star & $\rm{SNR_L}$ & $\rm{T_{eff}}$ & log(g) & $\rm{v_{micro}}$ & [Fe/H] & [Mg/Fe] & [Si/Fe] & [Ca/Fe] &... \\
\hline
2700.0 & 24 & 5010$\pm$85 & 0.69$\pm$0.42 & 1.73$\pm$0.13 & -1.88$\pm$0.14 & 0.29$\pm$0.15 & 0.57$\pm$0.18 & 0.48$\pm$0.20 & ...\\
9761.0 & 41 & 5138$\pm$71 & 2.22$\pm$0.02 & 1.27$\pm$0.07 & -0.82$\pm$0.11 & 0.38$\pm$0.15 & 0.19$\pm$0.18 & 0.22$\pm$0.20 & ...\\
11609.0 & 62 & 4333$\pm$142 & 1.39$\pm$0.22 & 1.47$\pm$0.07 & -0.72$\pm$0.11 & 0.51$\pm$0.15 & 0.25$\pm$0.18 & 0.37$\pm$0.20 & ...\\
7064.3 & 41 & 5127$\pm$87 & 2.37$\pm$0.02 & 1.69$\pm$0.10 & -0.79$\pm$0.15 & 0.31$\pm$0.15 & 0.34$\pm$0.18 & 0.21$\pm$0.20 & ...\\
697.0 & 53 & 4697$\pm$113 & 1.76$\pm$0.34 & 1.32$\pm$0.17 & -1.65$\pm$0.14 & 0.25$\pm$0.15 & 0.31$\pm$0.18 & 0.37$\pm$0.20 & ...\\
4953.1 & 35 & 5496$\pm$134 & 2.05$\pm$0.25 & 1.60$\pm$0.21 & -1.16$\pm$0.22 & 0.43$\pm$0.15 & 0.34$\pm$0.18 & 0.30$\pm$0.20 & ...\\
4239.1 & 14 & 5033$\pm$188 & 2.90$\pm$0.56 & 1.00$\pm$0.13 & -0.99$\pm$0.17 & 0.41$\pm$0.17 & 0.29$\pm$0.20 & 0.28$\pm$0.20 & ...\\
2860.0 & 91 & 4569$\pm$65 & 0.99$\pm$0.27 & 1.55$\pm$0.12 & -2.30$\pm$0.15 & 0.60$\pm$0.12 & 0.58$\pm$0.18 & 0.63$\pm$0.20 & ...\\
6577.0 & 40 & 4400$\pm$193 & 1.62$\pm$0.56 & 1.42$\pm$0.08 & -0.64$\pm$0.12 & 0.49$\pm$0.15 & 0.16$\pm$0.18 & 0.44$\pm$0.20 & ...\\
5126.3 & 43 & 4784$\pm$63 & 2.27$\pm$0.38 & 1.22$\pm$0.06 & -0.86$\pm$0.10 & 0.32$\pm$0.12 & 0.22$\pm$0.18 & 0.28$\pm$0.20 & ...\\
3083.0 & 54 & 4778$\pm$66 & 1.87$\pm$0.30 & 1.79$\pm$0.11 & -1.30$\pm$0.19 & 0.17$\pm$0.15 & 0.19$\pm$0.18 & 0.23$\pm$0.20 & ...\\
1490.0 & 55 & 5234$\pm$9 & 2.59$\pm$0.45 & 1.38$\pm$0.07 & -0.80$\pm$0.10 & 0.39$\pm$0.15 & 0.28$\pm$0.18 & 0.21$\pm$0.20 & ...\\
42011.0 & 40 & 4559$\pm$19 & 0.98$\pm$0.63 & 1.45$\pm$0.14 & -2.65$\pm$0.19 & 0.40$\pm$0.11 & 0.60$\pm$0.22 & 0.42$\pm$0.20 & ...\\
1670.0 & 43 & 4920$\pm$153 & 2.73$\pm$0.34 & 1.34$\pm$0.12 & -0.81$\pm$0.21 & 0.28$\pm$0.12 & 0.26$\pm$0.18 & 0.25$\pm$0.20 & ...\\
6805.0 & 24 & 4373$\pm$547 & 1.38$\pm$1.15 & 1.71$\pm$0.26 & -0.55$\pm$0.37 & 0.48$\pm$0.15 & -0.04$\pm$0.18 & 0.51$\pm$0.20 & ...\\
1697.2 & 22 & 5001$\pm$52 & 2.18$\pm$0.26 & 1.57$\pm$0.11 & -0.85$\pm$0.15 & 0.22$\pm$0.15 & 0.31$\pm$0.18 & 0.28$\pm$0.20 & ...\\
5953.0 & 15 & 4243$\pm$367 & 1.38$\pm$1.22 & 1.46$\pm$0.21 & -0.40$\pm$0.25 & 0.46$\pm$0.15 & -0.06$\pm$0.18 & ...$\pm$... & ...\\
12931.0 & 49 & 4612$\pm$81 & 0.88$\pm$0.08 & 1.34$\pm$0.10 & -3.20$\pm$0.10 & 0.55$\pm$0.11 & 0.71$\pm$0.18 & 0.58$\pm$0.20 & ...\\
6531.3 & 31 & 4751$\pm$285 & 1.96$\pm$0.30 & 1.99$\pm$0.16 & -0.98$\pm$0.28 & 0.05$\pm$0.15 & 0.24$\pm$0.18 & 0.64$\pm$0.20 & ...\\
9094.0 & 49 & 4644$\pm$261 & 0.78$\pm$0.66 & 2.50$\pm$0.26 & -2.31$\pm$0.27 & 0.49$\pm$0.12 & 0.58$\pm$0.18 & 0.55$\pm$0.20 & ...\\
1067.0 & 34 & 4372$\pm$153 & 2.04$\pm$0.47 & 1.39$\pm$0.13 & -0.47$\pm$0.19 & ...$\pm$... & 0.22$\pm$0.18 & 0.29$\pm$0.20 & ...\\
7604.0 & 29 & 4754$\pm$110 & 1.65$\pm$0.61 & 1.20$\pm$0.08 & -1.44$\pm$0.10 & 0.22$\pm$0.15 & 0.16$\pm$0.18 & 0.32$\pm$0.20 & ...\\
25782.0 & 35 & 4734$\pm$52 & 1.36$\pm$0.61 & 1.49$\pm$0.11 & -2.57$\pm$0.15 & 0.49$\pm$0.11 & 0.72$\pm$0.18 & 0.62$\pm$0.20 & ...\\
6373.1 & 23 & 5291$\pm$6 & 2.44$\pm$1.20 & 1.37$\pm$0.13 & -1.11$\pm$0.12 & 0.40$\pm$0.15 & 0.41$\pm$0.19 & 0.23$\pm$0.20 & ...\\
6382.0 & 46 & 4517$\pm$83 & 1.69$\pm$0.15 & 1.09$\pm$0.05 & -0.82$\pm$0.09 & 0.33$\pm$0.15 & 0.28$\pm$0.18 & 0.35$\pm$0.20 & ...\\
644.0 & 129 & 4768$\pm$16 & 1.54$\pm$0.07 & 1.45$\pm$0.07 & -1.57$\pm$0.06 & 0.41$\pm$0.15 & 0.40$\pm$0.18 & 0.45$\pm$0.20 & ...\\
\hline
\end{tabular}
\flushleft{A section of the online table with $\rm{SNR_L}$, \teff, \logg, [Fe/H], \vmicro\ and [X/Fe] for 22 elements for all 26 stars. The uncertainties in the elemental abundances are derived by adding the typical sensitivities on the stellar parameters and the internal error in quadrature }
\end{table*}

\begin{table*}

\caption{Line-By-Line abundances with atomic data and NLTE corrections where available.}
\label{tab:line}
\begin{tabular}{ccccccc}
\hline\hline
Star & Element & Wavelength & log($gf$) & $\rm{\chi}$ & $\rm{log(\epsilon)}$ & $\rm{\Delta NLTE}$ \\
& & (\AA) & & (eV) & \\
\hline
2700.0 & Mg II & 5711.1 & -1.724 & 4.346 & 6.07 & 0.17 \\
2700.0 & Si II & 5645.6 & -2.043 & 4.93 & 6.17 & -0.13 \\
2700.0 & Si II & 5665.6 & -1.94 & 4.92 & 6.29 & -0.13 \\
2700.0 & Si II & 5684.5 & -1.553 & 4.954 & 6.18 & -0.12 \\
2700.0 & Si II & 5948.5 & -1.13 & 5.082 & 6.22 & -0.03 \\
2700.0 & Si II & 6347.1 & 0.169 & 8.121 & 6.28 & ... \\
2700.0 & Ca II & 5260.4 & -1.719 & 2.521 & 5.31 & ... \\
2700.0 & Ca II & 5349.5 & -0.31 & 2.709 & 4.71 & ... \\
2700.0 & Ca II & 5582.0 & -0.555 & 2.523 & 5.00 & ... \\
2700.0 & Ca II & 5588.7 & 0.358 & 2.526 & 4.96 & ... \\
2700.0 & Ca II & 5590.1 & -0.571 & 2.521 & 4.72 & ... \\
2700.0 & Ca II & 5857.4 & -0.571 & 2.521 & 4.82 & ... \\
2700.0 & Ca II & 6102.7 & -0.85 & 1.879 & 5.02 & ... \\
2700.0 & Ca II & 6166.4 & -1.142 & 2.521 & 5.01 & ... \\
2700.0 & Ca II & 6169.0 & -0.797 & 2.523 & 4.92 & ... \\
2700.0 & Ca II & 6169.6 & -0.478 & 2.526 & 4.89 & ... \\
2700.0 & Ca II & 6439.1 & 0.39 & 2.526 & 5.17 & ... \\
2700.0 & Ca II & 6493.8 & -0.109 & 2.521 & 4.86 & ... \\
2700.0 & Ti II & 4865.6 & -2.7 & 1.116 & 3.01 & ... \\
2700.0 & Ti II & 4997.1 & -2.07 & 0.0 & 3.32 & 0.71 \\
2700.0 & Ti II & 4999.5 & 0.32 & 0.826 & 3.19 & 0.45 \\
2700.0 & Ti II & 5145.5 & -0.54 & 1.46 & 3.63 & ... \\
... & ... & ... & ... & ... & .. & ... \\
\hline
\end{tabular}
\flushleft{The star's id is given in column 1. Columns 2, 3, 4, and 5 give the element, wavelength, log($gf$), and excitation potential of the line, respectively. The derived abundance for the line is given in column 6 with the NLTE correction to that abundance given in the last column.}
\end{table*}

\section*{Acknowledgements}
{\small 
KH is partly supported by a Research Corporation Time Domain in Astronomy Grant. ML is partially supported by UT Austin College of Natural Science Fellowship. TB and SF were funded by the project grant ``The New Milky Way'' from the Knut and Alice Wallenberg Foundation. LC is the recipient of the ARC Future Fellowship FT160100402. KF is partly supported by ARC grant DP120104562. A.F.M. has received funding from the European Union's Horizon 2020
research and innovation programme under the Marie Sk\l{}odowska- Curie
Grant Agreement No. 797100. CK acknowledges support from the UK\textquotesingle s Science and Technology Facilities Council (grant ST/R000905/1). This research was partly supported by the Australian Research Council 
Centre of Excellence for All Sky Astrophysics in 3 Dimensions (ASTRO 3D), 
through project number CE170100013.

This work has made use of data from the European Space Agency (ESA)
mission {\it Gaia} (\url{https://www.cosmos.esa.int/gaia}), processed by
the {\it Gaia} Data Processing and Analysis Consortium (DPAC,
\url{https://www.cosmos.esa.int/web/gaia/dpac/consortium}). Funding
for the DPAC has been provided by national institutions, in particular
the institutions participating in the {\it Gaia} Multilateral Agreement.}

\bibliography{bibliography}
\bsp	
\label{lastpage}
\end{document}